# Hydrated Organic Matter: Functional Relationships with the Mass Physical Static and Dynamic Properties of Marine Mud

Richard H. Bennett,[1] Matthew H. Hulbert,[2] Roger W. Meredith [3]
[1&3] SEAPROBE Inc., 501 Pine St., Picayune, Mississippi 39466,
email: rhbenn_seaprobe@bellsouth.net
[2] Research Dynamics, 4001 Glacier Hills Drive, Unit 247, Ann Arbor, Michigan 48105,
email: mhulbert@researchdynamics.us

**ABSTRACT**

The mass physical and dynamic properties of marine mud deposits are a function of the interaction among a maximum of four sediment Phases as follows: (1) clay mineral solids that often include silt and sand size particles of different mineralogy, (2) semisolid hydrated organic matter, (3) free pore water, and (4) free gas when present. Historically, the total water was removed by oven drying which includes the seawater of hydration of the organic matter (OM); thus, the total water content was considered free pore water in several research disciplines. The semisolid hydrated OM resists transport through the mud pore space and the water that is bound to the OM is not free pore water. Calculations of marine mud mass-physical properties without consideration and correction for the presence of hydrated organic matter can introduce significant error in the amount of free water content of the pore space in the mud and the quantitative differences in both porosity and void ratio. Simulations reveal absolute relative differences due to the presence of hydrated OM that can reach meaningful values even when the %TOC values are below ~3%. Relevance is determined by comparing the differences due to the hydrated OM Phase with the differences in the fluid saturation Phase at 100% and at 90%. Understanding the presence and characteristics of hydrated OM is expected to engage important future research in several sciences and in geotechnical engineering of terrestrial and marine mud deposits.

## I. *INTRODUCTION*

This research addresses multi-disciplinary technical issues including quantitative and modeling of marine fine-grained sediments and the mass physical properties important in the analysis of mud deposits at large and small spatial scales. Major topics herein include quantitative analysis of sediment States and Phases, in addition to seafloor processes that drive the development and changes in marine mud multi-phase physical properties. Biogeochemical properties of marine mud deposits and the time-dependent processes (e.g., rates of deposition, organic degradation, and mud consolidation) alter sediment mass physical properties and the static and dynamic behavior and variability. The presence of hydrated organic matter, $OM_h$, primarily affects the sediment States and properties including (1) free pore water available in clay fabric pore space, (2) water volume of OM hydration, (3) reduction of permeability by volumetric contribution of $OM_h$ in pore space, (4) density reduction of dry OM by seawater hydration [Eq. (4) and (5)] dynamic behavior by physico-chemical attachment of $OM_h$ to clay particles in potential energy fields created by the clay fabric signatures. High percentages of $OM_h$ result in a higher degree of compressibility at considerably lower stresses than with little $OM_h$ and may reduce or eliminate acoustic shear wave propagation as a function of the organo-clay fabric and the physico-chemistry (clay microstructure) in mud deposits (Bennett and Hulbert, 1986).

This paper addresses issues concerning small scale properties that are important in understanding the behavior of acoustic energy propagation in marine mud deposits and specifically to acoustic shear wave energy propagation in marine muds containing hydrated organic matter.

## II. *SEDIMENT PHASES*

Marine mud consists of up to four types of material and each type is termed a sediment Phase. The physical and chemical properties of the Phases differ significantly from each other. The Phases are:

(1) siliciclastic fine-grained particles (dominantly clay minerals and often with different mineralogy) generally consisting predominantly of clay-size particles (≤3.9 µm) with silt-size particles (>3.9 µm - 62.5 µm) and often with some fine-grained sand-size particles

(2) hydrated organic matter which is not in solution and may include living microorganisms and their various metabolic products and can vary in density and degree of degradation (Evgenii A. Romankevich, 1984). Hydrated OM is a semisolid, flows less freely than free pore water and will be shown to significantly modify sediment properties

(3) seawater, with a density that is dependent upon salinity and temperature The quantity of OM dissolved in the seawater is typically orders of magnitude smaller than the semisolid hydrated OM.

(4) free gas which may or may not be present and with a density very dependent on pressure.

The fundamental sediment Phases and their mass-volumetric contributions are significant determinants that drive the static physical and dynamic geotechnical and geoacoustic properties of marine mud deposits under stress and strain conditions. Sediment Phases are the "building blocks" of the sediment States of marine mud. The quantitative contribution of mud Phases and their properties at nanometer to meter scales is important in many research areas. The research discussed here is multidisciplinary and complex in terms of the marine biogeochemical influence on the sediment properties. (Richards, 1967; Lambe and Whitman, 1969; Hulbert and Bennett, 1982; Bennett and Nelsen, 1983; Bennett and Hulbert, 1986; Bennett *et al.,* 1985, 1999a and b, 2012, 2013; Mitchell, 1993; Ransom, 1994; Hulbert *et al.,* 2002).

## III. *SEDIMENT STATES*

Fine-grained marine sediment States are defined as a function of specific empirical measurement techniques termed the Atterberg limits (Atterberg, 1911; Terzaghi, 1925; Casagrande and Fadum, 1940). The Atterberg limits provide the identification of the boundary limits and classification of cohesive sediment (Casagrande, 1948). The two Atterberg limits are the liquid limit and the plastic limit. Each term is well described by James K. Mitchell, 1993 as follows. “The liquid limit test is analogous to a dynamic shear test. Casagrande (1940 and 1932) deduced that the liquid corresponds approximately to a water content at which a soil has a shear strength of about 2.5 kPa.” “The plastic limit has been interpreted as the water content below which the physical properties of the water no longer correspond to those of free water (Terzaghi, 1925) and as the water content at which the cohesion between particles or groups of particles is sufficiently low to allow movement but sufficiently high to allow particles to maintain the molded positions (Yong and Warkentin, 1966).” When the Atterberg limits of a mud deposit reach the liquid and plastic limits (LL and PL) with depth in a mud deposit that essentially defines the boundary layer depth for the fluid and plastic mud deposits. See figure 1.

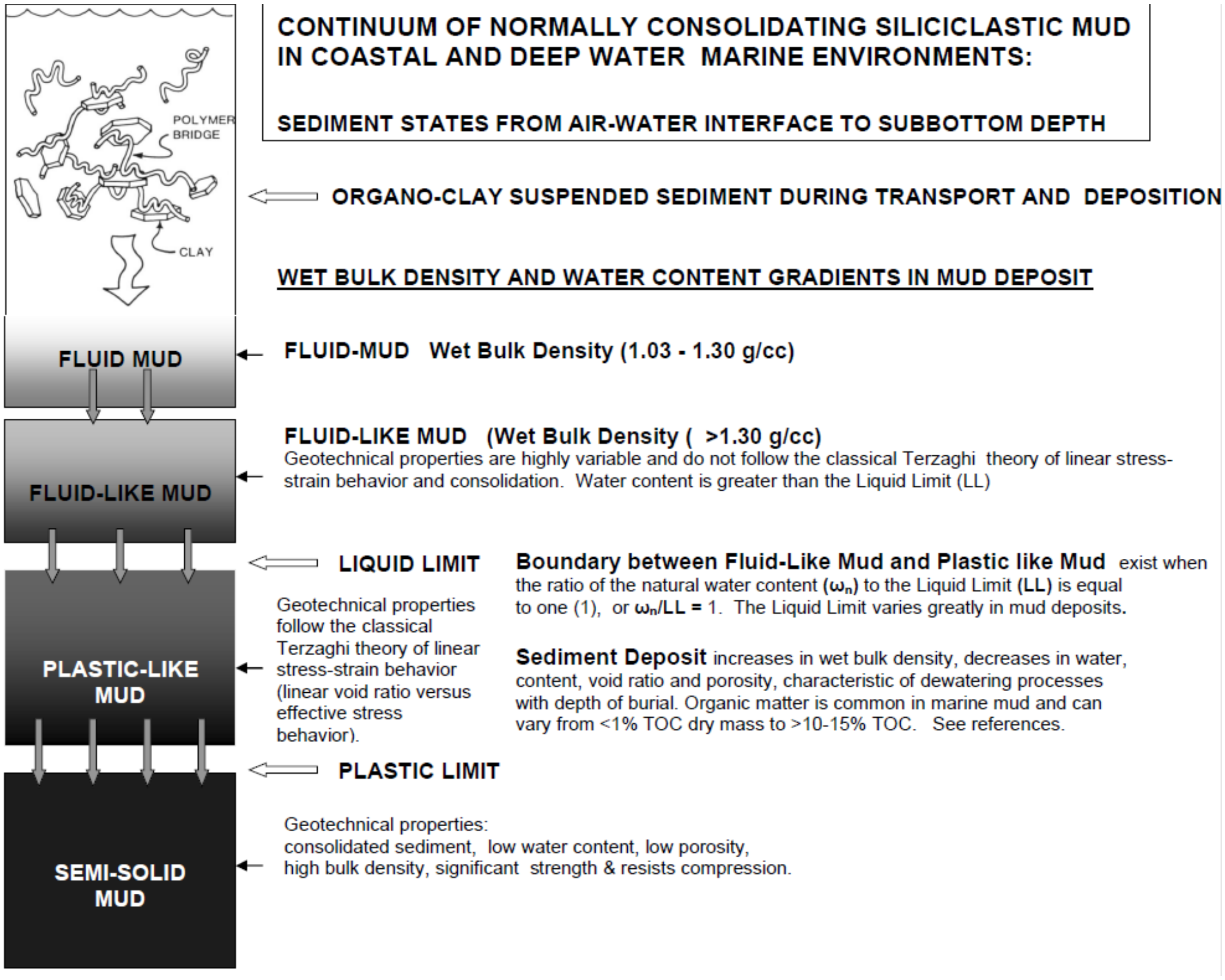

FIG. 1. Description of suspended sediment in the water column and the sediment States from the sediment-water interface to the semi-solid mud typical of normal consolidated marine muds. The sequence of sediment States is defined by the mass physical properties with subbottom depth. For detailed sediment properties see: Terzaghi and Peck(1967): Lambe and Whitman (1969); and Mitchell (1993).

The sediment States express the degree of consolidation and rigidity, a decrease in water content with depth, and significant changes in the Atterberg limits and change in the State. The Atterberg limits have been used effectively for nearly a century by various disciplines. In normally consolidated mud, the sediment States are defined and characterized from the sediment-water interface to depth as (1) fluid mud, (2) fluid-like mud, (3) plastic-like mud, and (4) semisolid mud. The properties and characteristics of each sediment State are defined descriptively in Figure 1 and details of the measurement techniques can be found in the literature cited herein.

Fluid muds of interest here, are the coastal, riverine, and offshore deep water deposits, and are especially important in tidal dominated depositional environments Fluid muds have an incredibly low wet bulk density defined on the basis of density (1.03 - 1.30 g/cc, Fig. 1). This often results in a deposit easily eroded from the site of deposition and which can be transported great distances in rivers and marine hydro-dynamically active environments.

Fluid muds are the most recent deposits at the sediment-water interface. In geological environments of sediment deposition over long periods of time, the muds gradually dewater and become higher density fluid-like muds >1.3 g/cc. With continual sedimentation and consolidation, fluid muds dewater to plastic-like deposits with increasing depth below the sea floor. Knowing the developmental history and the surficial to subbottom properties of marine and terrestrial sedimentary deposits that include the quantitative contributions of the fluid mud is crucial in providing a comprehensive understanding of the geophysical properties (e.g., the differences and variability in acoustics wave speed and shear speed), and the geological and geotechnical properties (mass physical and mechanical) of mud deposits.

## IV. *CONSOLIDATION PROCESSES AND PROPERTIES WITH $OM_h$, EXAMPLE*

An example of a normally consolidated marine mud deposit that includes siliciclastic minerals and $OM_h$ is depicted in Figure 2. These mud deposits exhibit a classic example of a curvilinear decrease in water content with depth below the seafloor, due to increasing overburden pressure. Multi-plate clay particles (domains) are morphologically "sheet-like" and carry significant electro-static charge in seawater and the $OM_h$ can attach to the faces and the edges of domains. This particular classic mud deposit is located where the Patuxent River enters the Chesapeake Bay, MD. This data (Fig. 2) reveals surficial mud with high water contents characteristic of fluid mud in the upper 55-60 cm subbottom from the sediment-water interface and fluid-like mud to depths of ~210 cm where the consistency and rigidity of the deposit changes to a plastic-like mud and the ratio of the total water content ($W_n$) to the liquid limit (LL) is ≤1.0 ($W_n$/LL ≤1.0). Surficial sediment in the Chesapeake Bay has been reported as high as 10% TOC (total organic carbon) and the average TOC in the Bay deposits is 2.2% TOC (percent dry mass, Kerhin *et al.,* 1988; Hill and Halka, 1988). Samples were collected at the sediment-water interface to ~5 cm subbottom and 10.0% TOC was measured for the mud characteristic of the mass physical properties depicted in Fig. 2 (Bennett *et al.,* 1995). Water content in Figure 2 is reported as percent dry mass. Historically and in this example as well, no corrections were made for hydrated OM to correct for the water of hydration and the mass of free seawater ($M_{fsw}$).

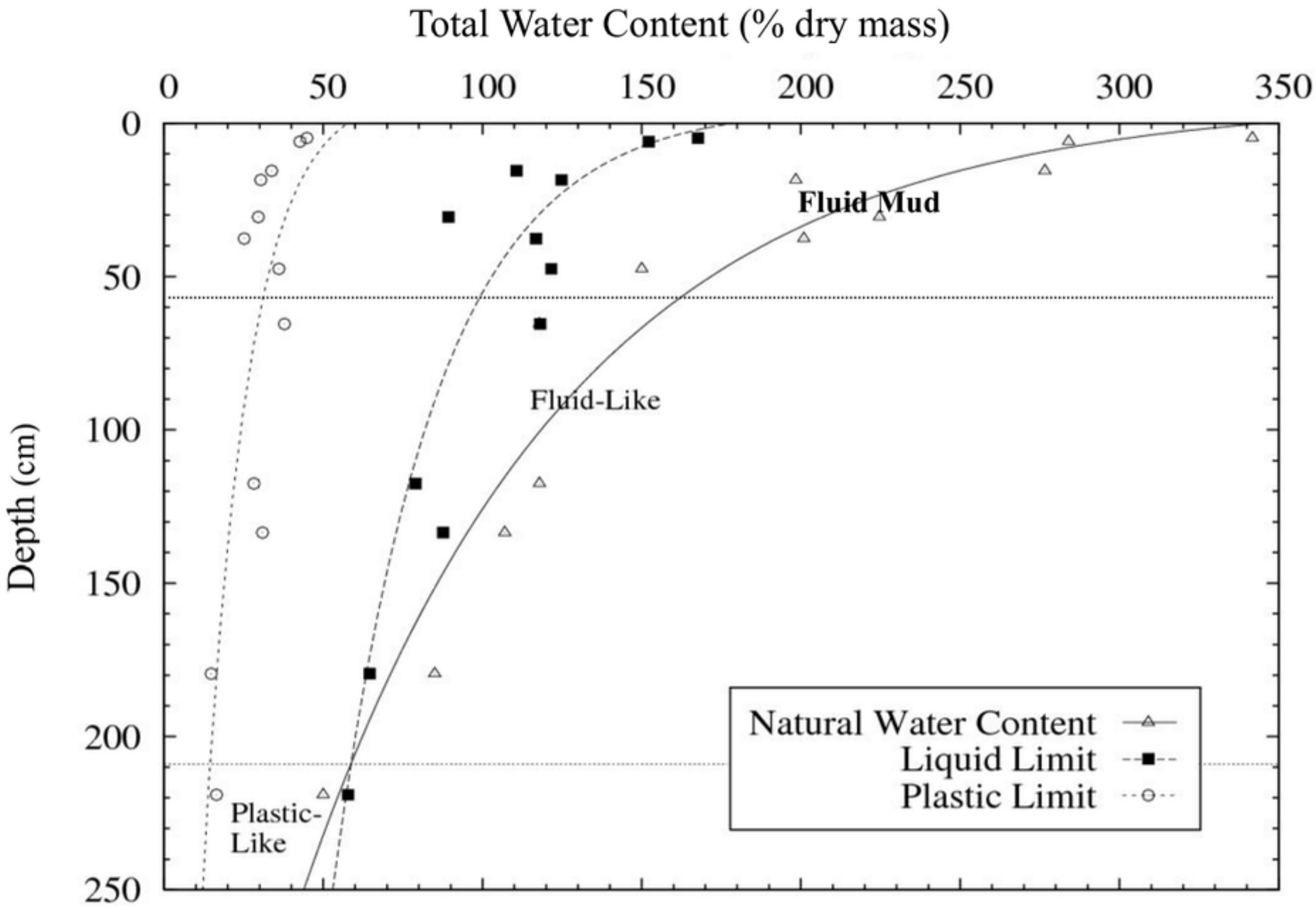

FIG. 2. A classical example of normally consolidated mud-state sequences (deposits) is typical of coastal embayment depositional environments. Both States and sequences may vary in depth, thickness, and water content. Note the high water contents characteristic of fluid mud to subbottom depths of ~50-60 cm and Fluid -Like mud to depths of ~210 cm where the water content is equal to the liquid limit (LL) and the mud deposit below that depth becomes plastic-like in stress-strain behavior. Sediment core data are from Chesapeake Bay near the area where the Patuxent River enters the Bay (Bennett *et al.,* 1995). Figure modified after Bennett *et al.,* 2011.

Not all marine mud deposits have the same stratigraphic sequences and changes in the mass physical properties and sediment States with depth of burial as depicted in Figures 1 and 2 that result in an increase in wet bulk density and decreases in porosity and void ratio with depth. Some deposits undergo processes that have stratigraphic States removed by erosion and/or marine slumping and thus formerly overlying sedimentary material is moved to another location (McGregor and Bennett, 1977, 1981; Bohlke and Bennett, 1980). When erosion or slumping occurs, the remaining deposit at the seafloor is exposed to the water column and to current activity including sediment deposition at the "new" sediment-water interface. This remaining deposit may consist of a more consolidated deposit than the material removed. In areas where slumping occurs, the seafloor at or near the sediment-water interface may be a shear zone formed during slumping. Properties of the shear zone have been found to differ from the typical earlier deposited high-water-content sediment. High-water-content mud may be deposited onto the shear zone following the slumping process, thus forming a more recent sediment-water interface.

The geotechnical (mass physical) properties of marine sediment deposits are highly variable in three dimensions at a wide range of scales (e.g., nanometers to meters) as a function of material types and processes. The variability in mud deposits is largely a function of: (1) different percentages of the material composition in the Phases present in the deposit, (2) the active processes such as static overburden stress and consolidation, dynamic stress-strain forcing by hydrodynamic wave activity, and shearing on slope (3) rates of sediment deposition and erosion, (4) changes in the clay microstructure with increasing overburden pressure with subbottom depth, (5) variability in clay minerology of deposits and in particle sizes, and (6) the biogeochemical processes active in marine mud such as degradation, excretion, and burrowing, and possible production of free gas.

## V. *CURRENT CLAY FABRIC MODELS OF MARINE MUD*

TEM imagery reveals and confirms that single clay particles are initially formed from several unit cells made predominately of two-to-one or one-to-one octahedral and tetrahedral sheets at the atomic level depending upon mineralogy (Chiou *et al.,* 2009). Domains are multi-plate face-to-face clay particles and are commonly the building-blocks of clay fabric. Clay domains attach to adjacent domains in the water column and at the sediment-water interface during deposition. The domains can attach in only three possible contact geometries (or Signatures): edge-to-face (EF), edge-to-edge (EE) and face-to-face (FF) (Bennett and Hulbert, 1986). Depending upon the number of clay mineral particles available in close proximity, these contact geometries can form aggregates with during suspended sediment transport and deposition (Fig. 3). In general, microbes are present in mud deposits. Figure 4(a). Marine mud deposits often contain variable amounts of silt and sand size particles and, when the clay particles dominate volumetrically, the larger particles (sand and silt) virtually "float" in the clay matrix without contact to other large size particles.

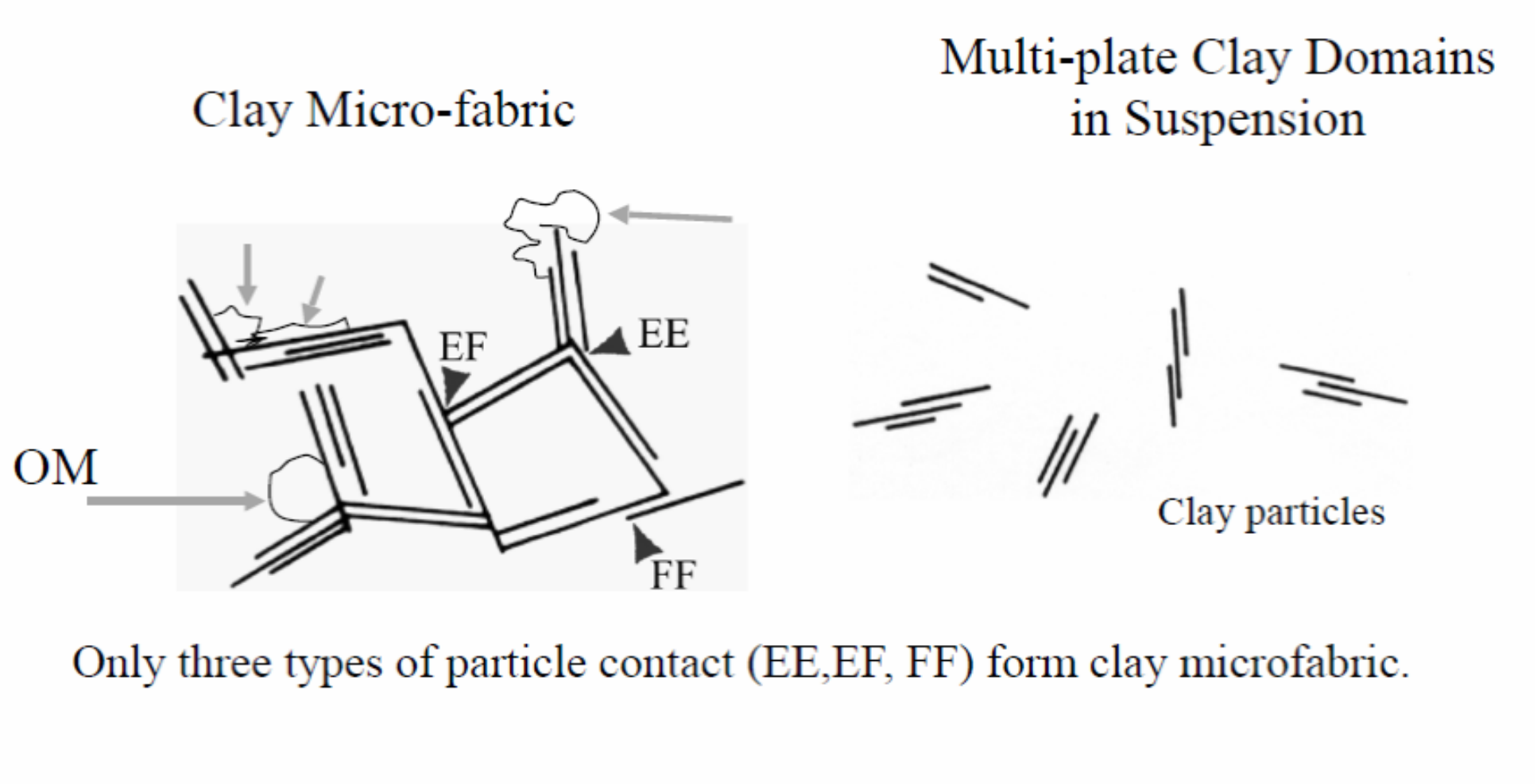

FIG. 3. Domains are layers of stepped and face-to-face multi-plate clay minerals (Moon, 1972). Three types of domain-particle contact (EF, FF, EE Signatures, Bennett and Hulbert, 1986) that form clay micro-fabric in the water column and at the sediment-water interface that become part of a marine mud deposit along with non-flocculated domains that settle at the sediment-water interface. Hydrated organic matter ($OM_h$) can attach to domains at particle contacts (e.g., EF), at edges of domains, on the faces of domains, and also interpose at multiple domain edges. Examples: See arrows pointing to OM attached to clay particles in Fig. 3.

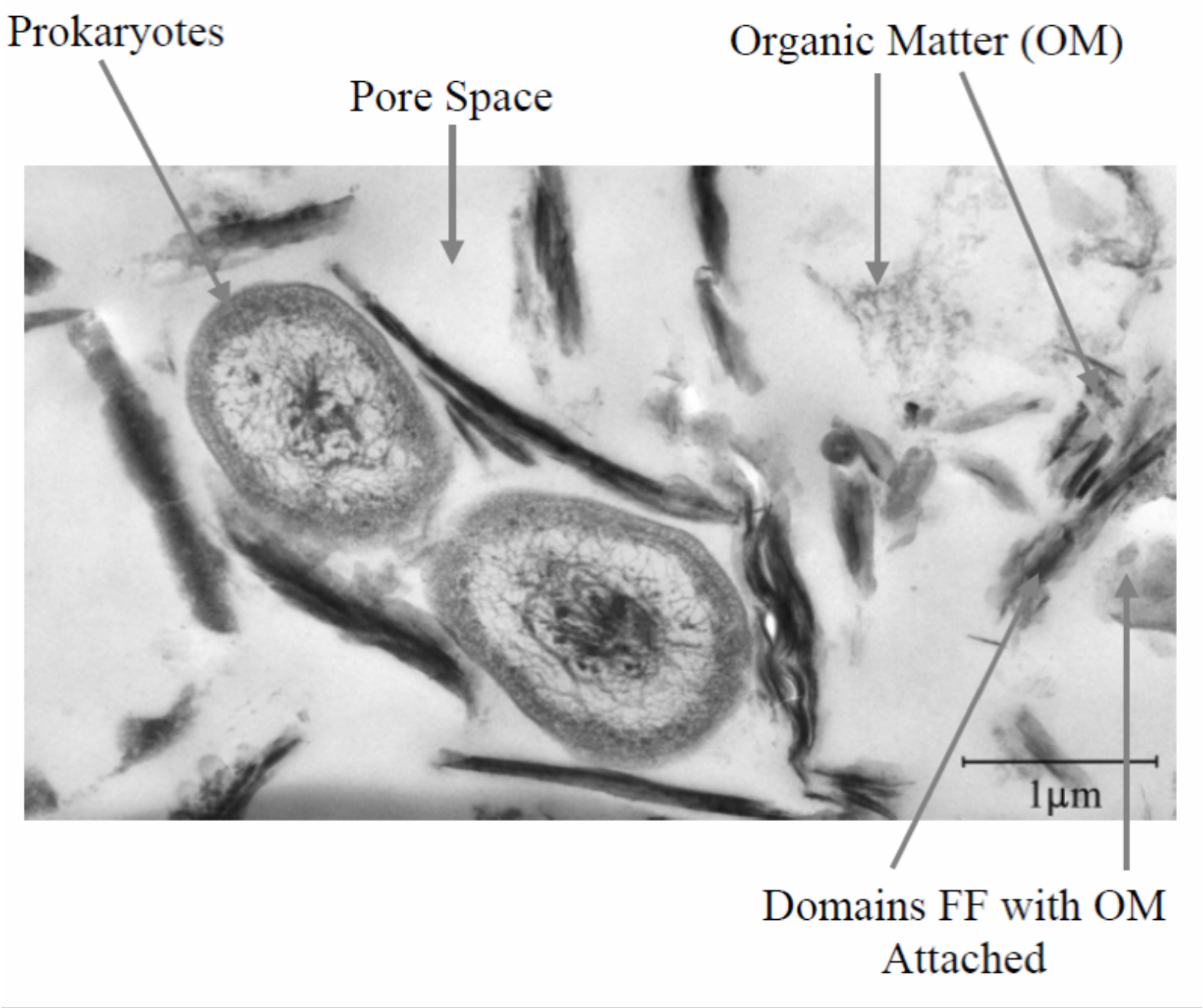

FIG. 4(a). TEM photomicrograph of two prokaryotes (one cell organisms) with domains (multi-plate clay particles) forming an "onion skin" lining around the bacteria, collected from 3 cm below the sediment-water interface in 720 m of water from box core 33 subsampled from the California continental slope South Mendocino Transect. Note large open pore spaces and $OM_h$ in pore space and attached to clay domains. Modified from Bennett *et al.,* 2004 SEPM-AAPG Presentation and Paper.

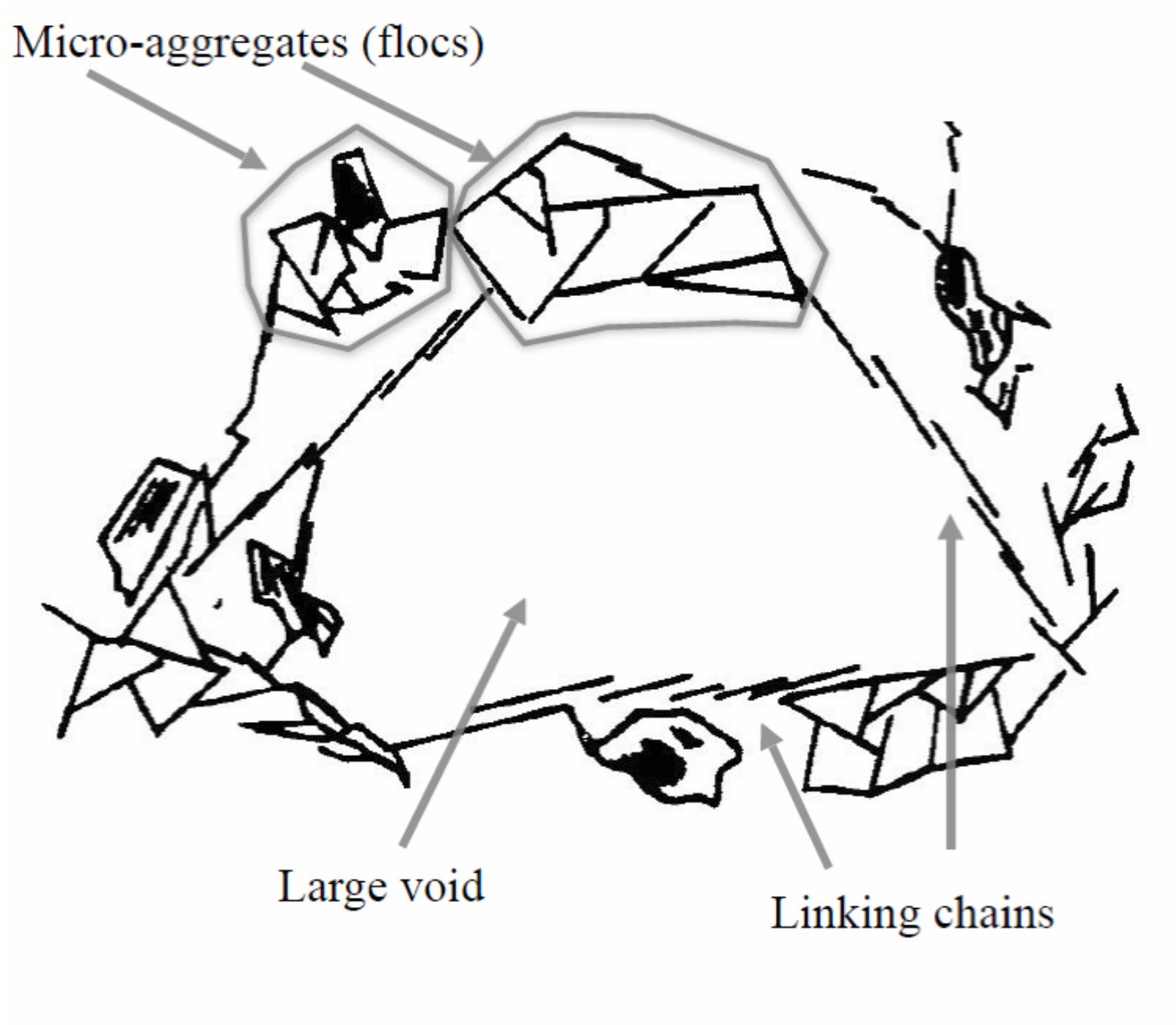

FIG. 4(b). Model of Red Clay observed in TEM images from the Pacific Ocean, modified from Bennett *et al.,* 1977. Note long face-to-face domain (thin) particles forming linking chains between aggregates. Compare with Fig. 4(c), image of Red Clay from the Pacific Ocean.

***A. The Marine Mud Clay Fabric Models consist of the following***

i. Made up of mineral components of various sizes and shapes including a significant proportion of small clay mineral platelets, organic matter, and detritus in a watery matrix.

ii. Clay mineral platelets are made up of variable numbers of layers, and thus of variable thickness, all bonded together in a sheet-like crystalline lattice.

iii. Clay mineral platelets typically are firmly attached to other platelets in variable numbers forming “domains” of variable thickness; this attachment is face-to-face, and the domain’s structure is a crystal with multiple severe dislocations/intrusions at the planes between individual platelets.

iv. In the presence of seawater, particles bond to each other with the clay domains and organic matter interacting via electrical charges on the faces and along edges of the domains and with the electronic structure of $OM_h$. The electrical charges generate forces that are comparable to and can be slightly larger than the gravitational force acting on these small domains in the surficial sediment (Bennett *et al.,* 1996).

v. Domains attach with others in open arrays, with voids filled with water and often with appreciable amounts of hydrated organic matter and sometimes free gas. The domains are generally randomly oriented at the sediment-water interface and in the surficial deposit with high water contents.

Some marine mud deposits in both shallow and deep water contain numerous aggregates of domains and $OM_h$. The domains of these aggregates are packed tightly together and have sufficient cohesion that the aggregates are distinct from the surrounding bulk clay fabric sediment in which domains form a more open array with larger pore spaces. Aggregates form during current flow by the combination of $OM_h$ and suspended and eroded mud fragments. When mud fragments include small aggregates, larger compound aggregates may be formed [Bennett *et al.*, 2013 and GCSSEPM 2012; Figure 4(d)].

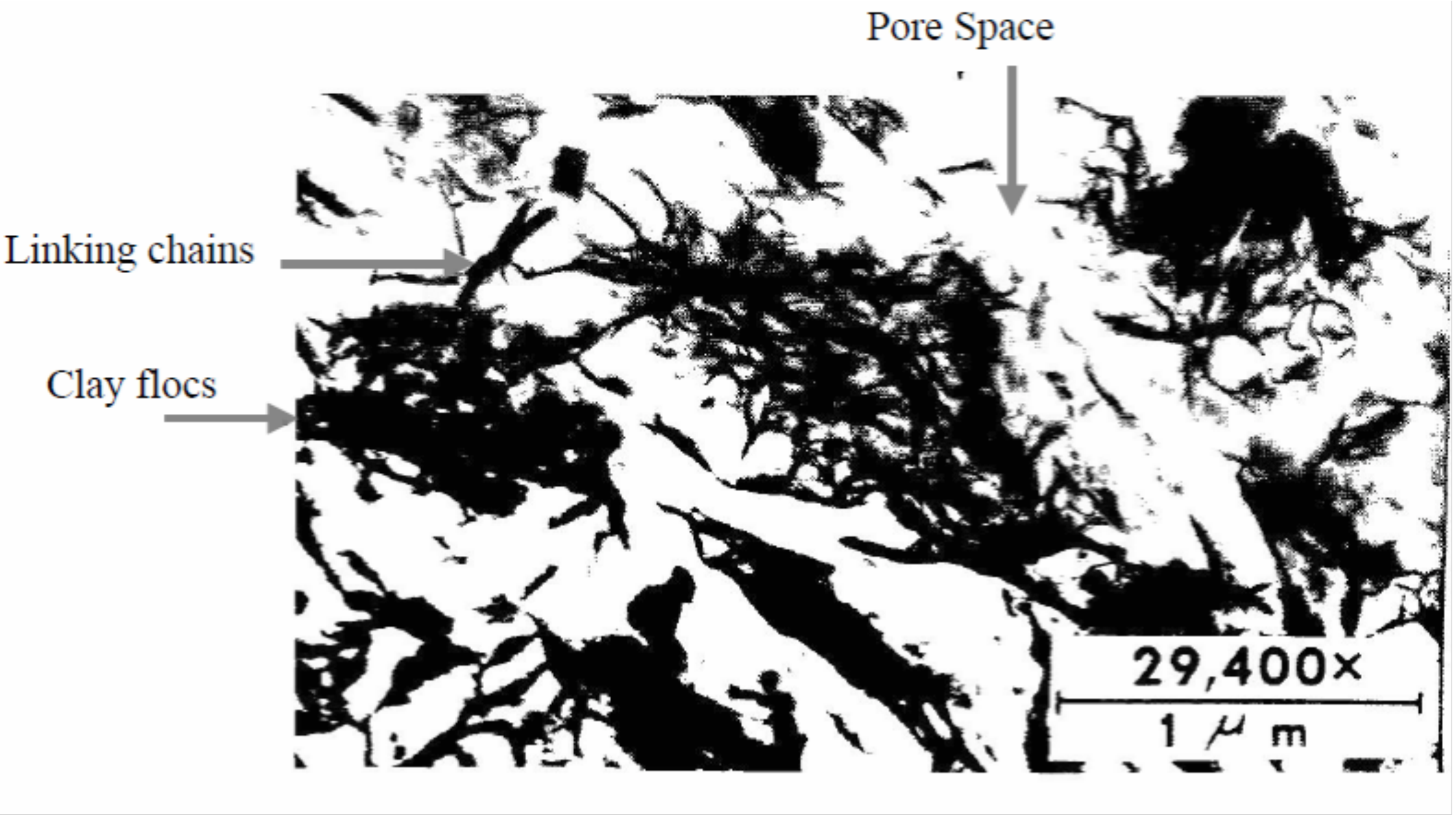

FIG. 4(c). Transmission Electron Micrograph of Red Clay from the Pacific Ocean. Note the aggregates with numerous small pores and much larger pores between aggregates. Observe the linking chains of domains, (many E-E) that connect aggregates. (Modified from Bennett *et al.,* 1977, 1981).

Unlike the pores of the bulk sediment, the numerous pores of the clay aggregates are small and poorly connected. As a result, water within these pores is unable to move as freely as in the bulk sediment surrounding the aggregate and, like water of OM hydration, is not part of the free pore water available for dynamic flow [see Fig. 4(a) and Fig. 4(c)]. Thus, measurements of permeability in marine mud containing $OM_h$ and clay aggregates involves initially, water in fluid flow ($W_{ff}$) = $W_t$ - ($W_{OMh}$ + $W_{CA}$) where:

$W_{ff}$ is the seawater available for flow driven by a pressure differential

$W_t$ is the total water in a unit volume of mud

$W_{OMh}$ is the hydration water bound in organic matter ($OM_h$)

$W_{CA}$ is the water constrained in the pores of aggregates

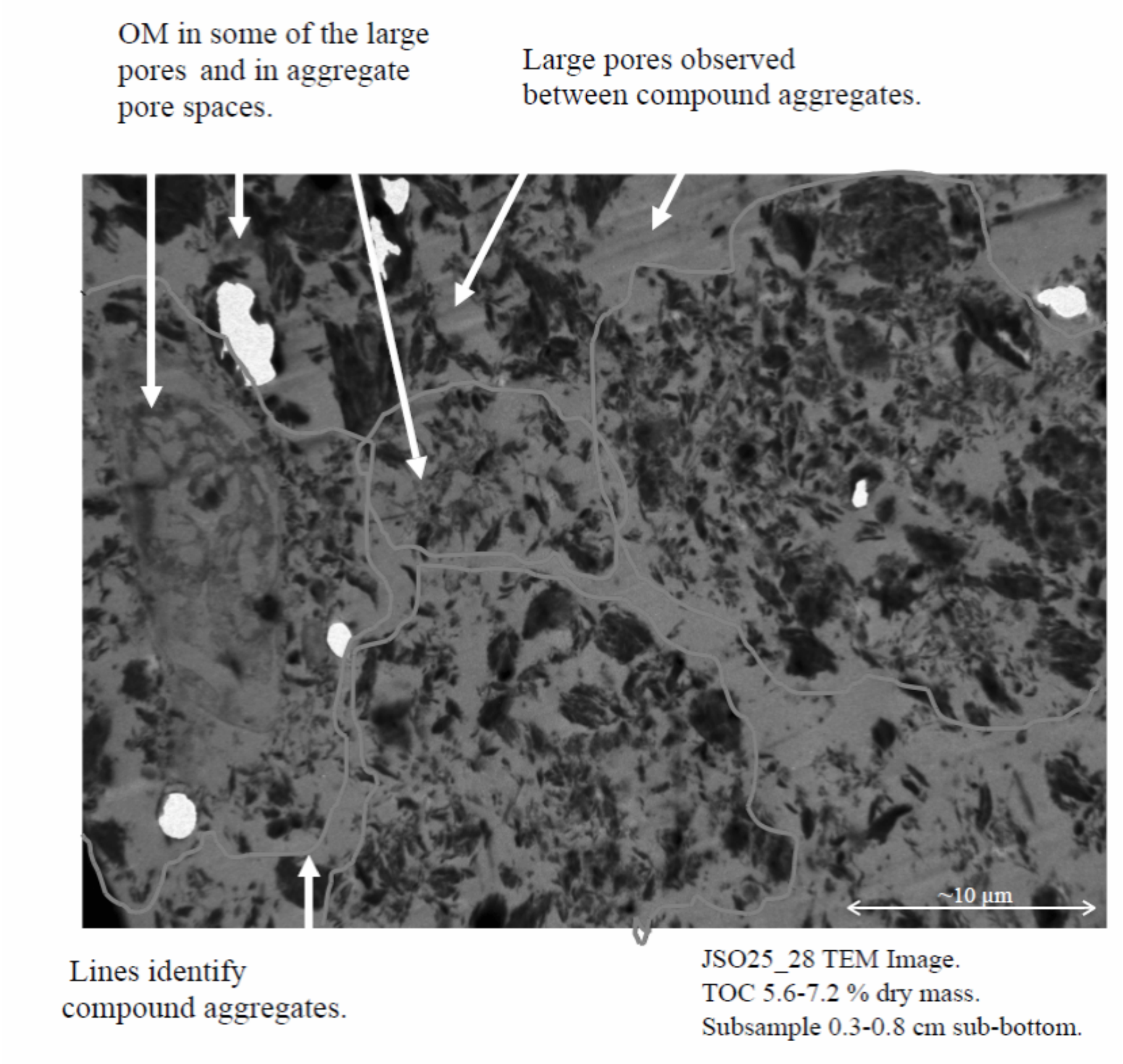

FIG. 4(d). Lines circumscribe compound aggregates that consist of previously-formed small aggregates, modified from Bennett *et al.,* 2012. Aggregates contain considerable organic matter with clay domains and the largest pores are found between compound aggregates. Numerous TEM images were made of subsamples collected from mud deposits of clay and OM during several Flume Experiments performed under the direction of Juergen Schieber, IU. TEM image courtesy of Jessica Douglas, formerly USM. Clay fabric interpretations by R. Bennett SEAPROBE, Inc. The research project was supported by NSF.

vi. With increasing subbottom depth of burial and greater uniaxial overburden stress, consolidation proceeds as a result of weaker domain particle contacts breaking and the domains commence a process of reorienting toward a predominantly FF arrangement resulting in mud with considerably high wet bulk density and very low porosity (Bennett *et al.,* 1977, 1981).

vii. Some marine mud deposits in shallow and deep water and on shallow or steep slopes develop dynamic stress conditions (e.g., from wave action or seismic vibrations) that drive the mud deposit down slope. This bulk motion develops a sub-\bottom shear zone with highly oriented face-to-face clay fabric signatures thus leaving a "new" sediment water interface (Bohlke and Bennett, 1980).

viii. Organic matter sorbs to the domains [Fig. 4] altering the bulk mechanical and physical properties of the mud deposit. The organic matter can be visualized in TEM imaging by staining the samples with lead and uranium compounds (Curry *et al.,* 2007)

## VI. *Analysis of Marine Mud*

Geoacoustic, geotechnical, geological, and biogeochemical research in marine mud requires exceptional care to obtain reliable measurement when using subsampling techniques and analyses to understand the complexity and quantitative variability of the sediment Phases, States, and physical and chemical properties. Great care is required in handling and transport of sediment cores to laboratories, and sometimes subsampling techniques and chemical treatment aboard ship are essential to preserve, as far as possible, the original (in situ) integrity of samples collected from the seafloor. Microbial activity degrades hydrated OM rapidly when exposed to temperatures more than the in situ mud sample temperature and samples must be kept above freezing to avoid disruption by ice crystals. Again, strict careful collection (coring) and sampling techniques including proper handling, storage, shipment from ship-to-shore-based laboratories, and laboratory techniques are required to obtain high quality samples and reliable data for modeling and analysis. Details of sediment sampling and techniques will not be addressed in this paper. A very substantial body of knowledge regarding appropriate procedures and techniques has been developed and published in peer-reviewed literature. Corrections and calculations for hydrated OM in clay mud should be considered in the analysis of the static and dynamic mass-physical properties.

### A. *Quantitative Treatment of Sediment Phases and Bulk Properties*

The following discussion focuses on the Phase properties of each marine mud subsample with the equations that are required to integrate component Phases to obtain corrected mass physical properties of the sediment States. Quantitative Phase properties include densities and specific volumes for each subsample representative of a selected sediment State. The mass physical properties include wet bulk density, water content, (percent dry mass and percent total mass), porosity, void ratio, percent saturation (and/or its complement, percent free gas). Void space in a sample is defined as the volume of the water plus the volume of any free gas present in a unit volume subsample. Details will be discussed, and equations are presented as required. Figures reveal the contrast between the corrected versus uncorrected data. When required, analytical techniques used in laboratories to obtain reliable data on each mud Phase and State will be mentioned briefly.

### B. *Siliciclastic Mineral Solids and Fluid Phases*

Both the mineral Phase, siliciclastic fine-grained particles (dominantly clay minerals), and the fluid Phase, seawater, have variable densities and specific volumes that are important in the calculations of their contributions to the sediment mass physical properties. These two Phases are fundamental contributors to the static and dynamic properties as well as the variability of marine mud deposits; these affect the response of not only acoustic energy but also the physical stress-strain behavior during hydrodynamic (wave) and mechanical loading of mud deposits. Variability exists within each Phase and significant variability is observed among the four Phases.

The various minerals commonly present in most marine mud deposits can exist in variable amounts (Table 1). An approximation of the total relative amount of each major fine-grained mineral type present, e.g. clay minerals, in a given volume of dry marine sediment can be determined in a laboratory by X-ray diffraction. Total mass-volume quantitative data per unit volume is questionable if corrections for each Phase are not made because of the presence of salts and any organic matter and/or gas. Standard laboratory procedures used to obtain water contents, porosity, wet bulk density, etc., of marine mud samples require oven drying, typically at an oven temperature of 105°C for 24 hours. This provides a quantitative measure of the total water content (mass and volume) when corrected for salinity (salt content) as discussed later in this section.

**MINERALS & SEAWATER COMMON IN MARINE MUD DEPOSITS**

| Seawater (Fluid Phase) | Density | Specific Volume | Temperature (of importance) |
|---|---|---|---|
| Seawater (35 ‰) | 1.0249 g/cc | 0.9757 cc/g | 20° C or 68° F |
| | 1.0282 g/cc | 0.9726 cc/g | 0.0° C or 32° F |
| Seawater (10 ‰) | 1.0058 g/cc | 0.9942 cc/g | 20° C or 68° F |
| | 1.0080 g/cc | 0.9921 cc/g | 0.0° C or 32° F |

| Mineral (Solid Phase) | Density(g/cc) | Specific Volume(cc/g) | Chemical Composition |
|---|---|---|---|
| Quartz | 2.65 | 0.3774 | $SiO_2$ |
| Albite Plagioclase | 2.60 | 0.3846 | $NaAlSi_3O_8$ |
| Kaolinite | 2.61 - 2.68 | 0.3831 - 0.3731 | $Al_2Si_2O_5(OH)_4$ |
| Chlorite | 2.60 - 3.50 | 0.3846 - 0.2857 | $(Mg_5Al)\ (Al\ Si_3)O_{10}(OH)$ |
| Mica Group Illite - aluminous some degraded | | | |
| 1. non-expandable | 2.76 - 3.00 | 0.3623 - 0.3333 | |
| 2. 10 Å layer silicates, aluminous some degraded | 2.66 - 2.69 | 0.3759 - 0.3717 | |
| Glauconite Fe(III)-rich non-expandable 10 Å layer silicate | 2.40 - 2.95 | 0.4166 - 0.3390 | $K_{0.75}(Fe_{1.75}Mg_{0.25})(Al_{0.5}0.5Si_{3.5})O_{10}(OH)_2$ |
| Smectite Group | 2.53 - 2.73 (low iron - high iron) | 0.3953 - 0.3650 | Na, K, ½$Ca_{0.3}\ (Al_{1.7}Mg_{0.3})Si_4O_{10}(OH)_2 \cdot nH_2O$ |

**TABLE I.** Common clay minerals and seawater properties found in marine and riverine mud deposits. Data from various sources including but not limited to: Grim, R. E., 1968, Ransom and Helgeson, 1994, Gaines, R. V., *et al.,* 1997, Rieder, M., *et al.,*1998. Note: the variable n in the Smectite group refers to the number of molecules of interlayer $H_2O$ that are associated with the cations in the interlayer of the mineral when hydrated as discussed by Ransom and Helgeson, 1994. For fully-hydrated smectites in seawater, n is ~4.5.

**C. *Organic Matter Phase***

Organic matter is commonly present in many marine mud deposits (Pennisi, 2020). When the total organic carbon content of the mud is at a level of 1% of the dry mass or more, essentially all the mass of OM is present as semi-solids and only an insignificant portion of the OM is dissolved in the pore water. Clay mineral surfaces adsorb limited amounts of seawater. In contrast, OM typically absorbs significant quantities of seawater thus is present in situ in the hydrated form, $OM_h$. These two Phases (clay minerals and $OM_h$) interact contemporaneously and have different stress-strain behavior under applied static and dynamic forces. Some, but not all, of the semisolid hydrated OM in marine mud is bound to the minerals. The proportion bound to the minerals is variable and may approach 100%. The strength of $OM_h$ binding to minerals is highly variable, ranging from very loosely to very tightly bound, even within a small mud sample. Mud in which all the pores are entirely filled with seawater and/or $OM_h$ (100% saturated) will behave differently to applied stresses than mud which has gas in some of the pore space. We note here that the $OM_h$ of living micro-organisms such as eukaryotic protists (e.g., amoebas) and of extracellular material such as mucopolysaccharide fibrils undoubtedly will affect the flux of water and other materials in the pore space (Bennett *et al.,* 1999b).

As observed in Figures 3 and 4(a), $OM_h$ can attach to each type of clay fabric signature and occupy pore space in the clay fabric structure. $OM_h$ bonds to clay fabric signatures in edge-to-face contact, on the flat faces, the edges and in the offset face-to-face contacts of clay particles with varying degrees of bonding strength (Bennett *et al.,* 1999a and b; Curry *et al.,* 2009; Bennett *et al.,* 2012). The charged or polar organic matter is bound to the charged clay mineral particles by physico-chemical interaction in seawater. Because $OM_h$ occupies pore space, it reduces the volume otherwise available to be occupied by freely flowing pore water. The process of $OM_h$ degradation can result in an increase in pore space available for pore water. Degradation of $OM_h$ may be retarded in mud deposits rich in smectite and illite where access to it by degrading enzymes is impeded by its being firmly bound by the deep physico-chemical energy fields at edge-to-face (EF) contacts (Curry *et al.,* 2007).

Research has revealed that, when a sufficient amount of OM (>2-3% TOC) is present in surficial mud, the result is high porosity and seawater content with low wet bulk density (Bennett *et al.,* 1985, 1999b; Busch and Keller, 1981, 1982). Figure 5 presents data from several environments characteristic of fine-grained deposits in coastal, deep water marine, bays, rivers and tributaries, etc., and provides the variability and ranges in TOC reported from numerous sources.

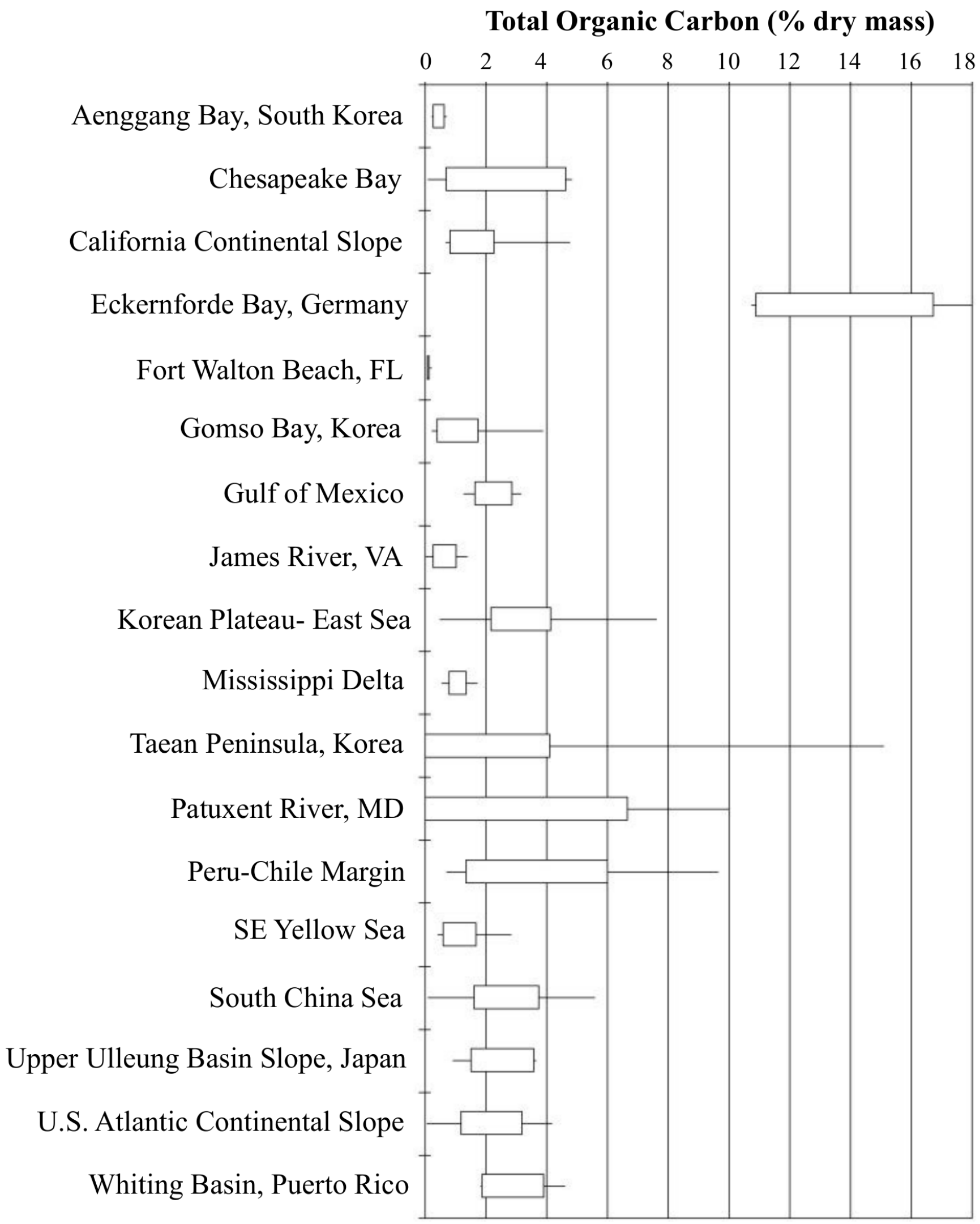

FIG. 5. Compilation was made of total organic carbon content, TOC % dry mass range ( +/- one std. dev. from average) from Mud Deposits in Various Seas, Bays, and Rivers. ONR funded research at many of these sites. (Compiled by C. Curry, SEAPROBE, Inc., 2003 and 2004).

**VII.** ***Mass -Volume Calculation and Determination of Marine Sediment Properties and Phases including 1) Pore Water and Salinity, 2) Hydrated Organic Matter, 3) Mineral Solids, and 4) Free Gas.***

The quantitative mass-volume contribution of each marine mud Phase present in sediment deposits determines the mass physical properties and supports the stress-stain behavior of marine mud when subjected to static and dynamic stresses. Hydrated organic matter in marine mud has been largely neglected in terms of quantifying its presence and role in the biogeochemical contribution to mass-physical properties, as well as its presence and impact on consolidation, stress-strain behavior, and acoustic propagation. Usually one of two methods are used to express the amount of organic matter in sediment: (1) The total amount of organic matter is rationalized in terms of the amount of TOC (total organic carbon) and is expressed as percent dry sediment mass (%TOC) when the sediment is dried for ~24 hours at ~105°C. (2). The amount of dry OM ($OM_d$) is calculated from the amount of TOC (see following section). When TOC is determined in the laboratory, procedures and techniques are used to correct for carbonates present in the sample. The mass-volume amount of a particular sediment sample depends upon the physical and chemical properties of each phase, degradation state of hydrated OM, and relative amounts of each Phase present and varies depending upon transport and post-deposition processes. Different marine mud deposits and sedimentary environments experience variable depositional rates, erosion, and times of deposition and various post-depositional diagenetic and biogeochemical processes including subbottom consolidation processes all of which ultimately determine the nature and sediment properties variability of a mud deposit. Sediment deposits are affected by externally and internally driven stress regimes (waves, currents, geologic stratigraphic forcing functions such as gravity) that are often strong and dynamic on steep and even on very gentle bathymetric slopes.

Each Phase provides not only essential information and data for simulating and predicting the multi-Phase material properties behavior under imposed and natural stresses, but also provides understanding of their role in addressing a wide range of science issues in marine research disciplines including acoustics, geotechnical engineering, geophysics, biogeochemical processes and marine sedimentology of fine-grained siliciclastic deposits. Given the four material Phases, calculations of the mass-volume property of each Phase is required. The following basic calculations apply not only to clay-mineral rich fine-grained marine mud but also directly to sandy-silty-clay fine-grained sediment.

## VIII. *CALCULATION OF SELECTED PROPERTIES: SEDIMENT PHASES*

### *A. Organic Matter Discussion*

Assume a measurement of TOC = 10% expressed as percent dry mass of sediment (corrected for the presence of carbonates). Applying the Redfield Ratio (Redfield *et al.,* 1962; Bennett *et al.,* 1999a and b), the TOC equals 37% of the dry marine OM mass. Therefore the mass in grams of: $OM_d$ = TOC/0.37 (or 2.7027 x TOC) multiplied by the grams of dry sediment following oven drying when TOC is expressed as g/g dry sediment [see Eq. (1) below]. The dry $OM_d$ density ($\gamma_{omd}$) is generally in the range of 1.40 $gcc^{-1}$ to 1.60 $gcc^{-1}$ and seldom has a value as high a 1.8 $gcc^{-1}$ (not considered here) (Tisa *et al.,* 1982; Feijo´Delgado *et al.,* 2013; Milo and Phillips, 2015; Schumacher, 2002). The reasonable average for dry $OM_d$ density ($\gamma_{omd}$) is 1.45 $gcc^{-1}$. The density ($\gamma_{omh}$) range for hydrated $OM_h$ is estimated to be from 1.10 $gcc^{-1}$ to 1.35 $gcc^{-1}$ (Tisa *et al.,* 1982; Milo and Phillips, 2015). The reasonable average for $OM_h$ density ($\gamma_{omh}$) is 1.22 $gcc^{-1}$.

***B. Water and Salt Discussion***

Salt-free Water Content ($\omega_n$), assumed = 200%, is expressed as percent dry mass of sediment. As discussed, marine mud may contain some silt size grains and sometimes also sand-size particles but is rich in clay minerals. This example uses an average grain density for the solids ($\gamma_{ms}$) of 2.68 $gcc^{-1}$. For simplicity of calculation we assume a dry mass of the sample to be 6.50 g (includes mass of solid minerals (CM), salt from pore water, and dry OM, each of which can be determined by laboratory measurement). At laboratory pressure the density of seawater is $\gamma_{sw}$ =1.0249 $gcc^{-1}$ at 20°C. The pore water salinity s = 35‰ and is expressed herein as parts per thousand (PPT) rather than Practical Salinity Units (PSU). The Initial Total Mass ($M_T$) of the subsample is determined and the mud sample is dried at 105°C for 24 hr (standard procedure) to determine the total amount of water salt free.

Example A. Given properties and parameters:

TOC = 10% dry mass (includes minerals, $OM_d$, and salt)

Marine $OM_d$ dry density ($\gamma_{omd}$) = 1.45 $gcc^{-1}$

Marine $OM_h$ hydrated density ($\gamma_{omh}$) = 1.22 $gcc^{-1}$ (includes saltwater in the $OM_h$, Water Content ($\omega_n$) = 200% as percent dry mass of sediment (includes minerals, $OM_d$, and salt) = uncorrected ($\omega_n$). Uncorrected here means there is salt in the dry marine sediment sample following drying at 105°C for 24 hours.

Pore water salinity (s) = 35‰

Mineral solids average grain density $\gamma_{ms}$ = 2.68 $gcc^{-1}$

Density of seawater ($\gamma_{sw}$) =1.0249 $gcc^{-1}$ at 20°C

The total dry mass of the laboratory sample ($M_s$) is assumed to be 6.50 g (includes: minerals (CM), organic matter ($OM_d$), and salt (dry). $M_{wsf}$ is the mass of the salt-free water driven off in drying at 105°C for 24 hours.

Initial Total Mass ($M_T$) of the marine mud subsample calculated after oven drying is:

$M_T = M_s + M_{wsf} =$ 6.5 g + (200% of 6.5) or 6.5 g + 13 g = 19.5 g Initial Total Mass

Example B. Organic Matter (OM). The gravimetric mass and volumetric contributions of $OM_d$ for any TOC value can be calculated by: TOC = 0.37 x $OM_d$ as percent dry mass (g) of carbon following drying at 105°C. This can vary somewhat depending upon type of $OM_d$ and degree of degradation. We use this as in the example computation.

Organic Matter $OM_d$ = TOC/0.37 or 2.7027 x TOC (1)

$M_s$ = 6.5 g = mass of dry sediment. Therefore, for a TOC = 10% of 6.5 g = 0.65 g of carbon and the dry organic matter mass is $M_{omd}$ = 2.7027 x 0.65 g = 1.7568 g of $OM_d$ without salt and without water. Given $M_{omd}$ for the bulk sample the volume of dry organic matter is $V_{omd}$ = 1.7568 gm/1.45 $gcc^{-1}$ = 1.2116 cc. The volume of $OM_h$ which has a density of 1.22 $gcc^{-1}$ is yet to be determined (See Example E below).

Example C. Seawater. Each gram of seawater including pore water salinity (s) of 35‰ has 0.035 g of sea salts. Therefore, the amount of water without salt in 1.0 g of seawater is $M_{wsf/g}$ = 1.000 – 0.035 = 0.965 gm.

Given in Example A above, uncorrected (salt-free) water content ($\omega_n$) = 200% is the amount of water driven off during drying and the salts remain in the dry sediment sample and 200% of the mass of the dry sediment with salt is water salt free ($M_{wsf}$).

water content $\omega_n = (M_{wsf}/M_s) \times 100 = 200\%$ where: (2)

$\omega_n$ = water content salt free

$M_{wsf}$ = mass of the salt-free water driven in drying at 105°C and

$M_s$ = mass of the oven dried solids remaining following drying (includes salt, $OM_d$, solid minerals (CM).

Given above, the dry sediment ($M_s$) = 6.5 gm

Therefore: using Eq. (2) for salt-free water content ($\omega_n$) = $M_{wsf}/M_s$ x 100 = 200%

or 2.00 = $M_{wsf}$/6.5 g and

$M_{wsf}$ = 2.00 x 6.5 g = 13.0 g of salt-free water in the laboratory sample.

For every gram of seawater there is 0.035 g of salt therefore:

Given 13.0 g of salt-free water the quantity of salt left behind in the dried laboratory sample is calculated according to:

0.965g = $M_{wsf/g}$ or the amount of salt-free water per gram of seawater at 35ppt

13.0 g/0.965 g x 0.035 g salt = $M_{st}$ = 0.4715 g salt

accordingly, 13.0 g + 0.4715 g = 13.4715 grams of seawater or $M_{sw}$ = 13.4715 g or

$M_{sw} = M_{wsf} + M_{st}$

The specific volume of seawater ($V_{sw}$) at 35‰ salinity (s) and at 20℃ is:

$V_{sw} = 1/\gamma_{sw}$ = 1/1.0249 $gcc^{-1}$ = 0.9757cc/g

Thus, the volume of the seawater ($V_{swc}$) corrected for salinity from subsample at laboratory temperature and pressure is: $V_{swc}$ = 0.9757cc/g x 13.4715 g = 13.1442 cc

A corrected saltwater content is $\omega_{nc} = (M_{wsf} + M_{st})/ (M_s - M_{st}) \times 100$ (3)

where $M_{sw}$ is the total mass of seawater in the sample. $M_{sw} = (M_{wsf}+M_{st})$

$\omega_{nc} = (13.00 + 0.4715)/ (6.5000 - 0.4715) = 13.4715/6.0285 = 2.23463 \times 100 = 223.46$

$\omega_{nc} = 223.46\%$ Dry Mass ($CM + OM_d$) corrected for Salinity

The in situ temperature is assumed to uniform throughout the collected sample.

Example D. Solid Mineral Phase. (clay <3.9 µm plus some silt <62.5 - ≥3.9 µm and sand ≥62.5 µm usually fine to medium size sand ~0.5mm as defined by the Wentworth 1922 size scale developed and still used).

The minerals in the laboratory sample are clays and have an average grain density = $\gamma_m$ = 2.68 $gcc^{-1}$, but we have a dry mass of 6.50 g ($M_s$) for the multi-component dry solids ($OM_d$, salts, and clay minerals). Therefore, the mass ($M_m$) of the clay minerals in the lab sample is:

$M_m = M_s - (M_{omd} + M_{st}) = 6.5$ g – (1.7568 g + 0.4715 g) = 6.5 g – 2.2283 g = 4.2717 g or

The mass of the clay mineral grains = $M_m$ = 4.2717 g

The average grain density of the mineral grains = $\gamma_m$ = 2.68 $gcc^{-1}$

The volume of the mineral grains = $V_{cm}$ = 4.2717 g/2.68 $gcc^{-1}$ = 1.5939 cc

Example E. Hydrated Organic Matter ($OM_h$). In the text example above using the Redfield ratio the dry mass ($M_{omd}$) of the organic matter was determined as:

$M_{omd}$ = 1.7568 g ($OM_d$ without salt and without water)

The $OM_d$ dry density is 1.45 $gcc^{-1}$ and the seawater hydrated OM density is 1.22 $gcc^{-1}$. In marine mud deposits hydrated organic matter is bonded physico-chemically with seawater and the hydration water will reduce the available free water pore volume and the porosity ($n_{cv}$).

In these calculations, we assume there is no change of total sample volume when organic matter is hydrated and that the salinity of the water of hydration is the same salinity as that of the free pore water. The organic matter will occupy pore space and is often observed, via Transmissions Electron Microscope (TEM) images, attached to clay particles (Avnimelech and Menzel, 1984; Avnimelech *et al.,* 1982; Baerwald *et al.,* 1991; Bennett *et al.,* 1991, 1996, 2012; Curry *et al.,* 2009). Just as seawater may be considered to be composed of dry salt and fresh water, the hydrated OM may be considered to be composed of $OM_d$ and bound seawater of hydration. Here we need to know what volume of seawater is required to hydrate $OM_d$ with a specific dry density $\gamma_{omd}$ of 1.45 $gcc^{-1}$ to achieve a specific hydrated organic matter ($OM_h$) density of $\gamma_{omh}$ = 1.22 $gcc^{-1}$. This can be determined from the definition of density for a two component mixture by solving for the volume of one component.

Fractional Component of Seawater, $Xf = (\gamma_{omh} - \gamma_{omd})/(\gamma_{sw} - \gamma_{omd})$ (4)

where: $\gamma_{omh}$ is the density of the hydrated OM ($OM_h$) =1.22 $gcc^{-1}$

$\gamma_{omd}$ is the density of the dry OM ($OM_{d)}$ =1.45 $gcc^{-1}$

$\gamma_{sw}$ is the density of Seawater =1.0249 $gcc^{-1}$

1. The calculation of the mass and volume contributions of $OM_h$ and seawater Phase in $OM_h$ having a density of 1.22 $gcc^{-1}$ is given by:

Xf= (1.22 - 1.45)/(1.0249 - 1.45) = (-0.2300)/(-0.4251) = 0.541049

1(a). Fractional Seawater volume; Xf = 0.541049

This volume Xf is the volume of seawater in 1 cc of $OM_h$ ($\gamma_{omh}$) at a 1.22 $gcc^{-1}$ density. The grams of SW having a volume of 0.541049 cc is:

1.0249 $gcc^{-1}$ x 0.541049 cc = 0.554521 g

1(b). Grams of SW = 0.554521 g (grams of seawater)

The volume of the $OM_h$: subtract Volume ($V_{sw}$) of the SW fraction [1(a)] in 1 cc of $OM_h$ having a density ($\gamma_{omh}$) of 1.22 $gcc^{-1}$ or (1.00000 cc - 0.541049 cc) = 0.458951 cc

1(c). $OM_h$ volume (0.458951 cc)

And the grams of $OM_d$ having a density of 1.45 $gcc^{-1}$ in 1.22 g of $OM_h$ is: 1.45 g/cc x 0.458951 cc = 0.665479 g

1(d). Grams of $OM_d$ = 0.665479 g

As a check on calculations:

The density of $OM_h$ ($\gamma_{omh}$) is: 0.665479 g $OM_d$ + 0.554521 g SW = 1.22 $gcc^{-1}$

The total grams of the dry $OM_d$ in the subsample determined above (Example B, salt free) is 1.7568 g with density ($\gamma_{omd}$) of 1.45 $gcc^{-1}$. The volume of the dry $OM_d$ ($V_{omd}$) in the subsamples is: ($V_{omd}$) = 1.7568 g/1.45 $gcc^{-1}$ = 1.211586 cc.

The total grams of seawater for 1.7568 g $OM_d$ (dry) of the $OM_h$ ($\gamma_{omh}$ =1.22 $gcc^{-1}$) is: (1.7568/0.665479) = 2.639903 then 2.639903 x 0.554521 = 1.463882 g seawater. See above calculations 1a - 1d for equation Xf and determination of grams of OM in B.

Then 1.463882 g SW + 1.7568 g $OM_d$ dry = 3.220682 g hydrated $OM_h$ in the mud subsample at the density of $\gamma_{omh}$ = 1.22 $gcc^{-1}$.

The total hydrated $OM_h$ volume ($Vom_h$) is the volume of the dry $OM_d$ ($V_{omd}$) plus the Volume of the Seawater of hydration ($V_{swh}$).

$V_{omd}$ = 1.7568 g/1.45 $gcc^{-1}$ dry = 1.211586 cc Dry OM

$V_{swh}$ = 1.463882 g SW/1.0249 $gcc^{-1}$ = 1.428317 cc Seawater of hydration

Thus, a total volume of 1.7568 g of seawater hydrated $OM_h$ ($Vom_h$) is:

$$Vom_h = V_{omd} + V_{swh} \quad (5)$$

$Vom_h$ = 1.211586 cc + 1.428317 cc = 2.639903 cc

As a check on calculations: (1.463882 g SW + 1.7568 g $OM_d$)/2.639903 cc = $\gamma_{omh}$ = 1.22000 $gcc^1$

## IX. *CORRECTED GRAVIMETRIC AND VOLUMETRIC QUANTITIES OF THE THREE PHASE MARINE MUD SAMPLE FOR 100% SATURATION*

### A. *Example Calculations of Sediment Phases*

Here we focus on a complex sediment type, marine mud, rich in clay minerals and organic matter with pore water salinity of 35% and initially 100% water saturated or without free gas (addressed later in paper). We start here with a set of laboratory data to provide quantitative examples of the calculations for each Phase present with no free gas in marine fine-grained sediment rich in organic matter and deposited in normal salinity water. Clay minerals are considered to be illite and smectite that are common constituents of marine mud deposits. The examples using corrected subsample mass and volume data for the three Phases comprising a mud subsample, assuming 100% seawater saturation, provide sediment mass physical properties data required for numerous types of technical analyses in marine acoustics and geology, geotechnique, and applied engineering geology, etc.

As mentioned above the dry mass of the subsample is 6.50 g (includes mass of solid minerals (CM), the dry organic matter (OM), and salt from pore water.

Total grams of water corrected for salt content = 13.4715 g SW

Total grams of dry solids CM +OM (without dry salt) = 6.0285 g CM +$OM_d$

Therefore: The total mass of the subsample is 13.4715 g SW + 6.0285 $OM_d$ and CM = 19.50 g

a. Organic Matter, OM

Density dry $\gamma_{omd}$ = 1.45 $gcc^{-1}$

Dry Mass = $M_{omd}$ = 1.7568 g

Seawater Mass of Hydration $M_{swh}$= 1.4639 g

Volume $OM_d$ dry = $V_{omd}$ = 1.2116 cc

Volume $OM_h$ Hydrated = $V_{omh}$ = 2.6399 cc

Total Mass $OM_h$ Hydrated = $M_{omh}$ = 3.2207 g

Density of Hydrated $OM_h$ $\gamma_{omh}$ = 1.22 $gcc^{-1}$

b. Seawater Mass -Volume and Salinity (S) = 35‰

Density = $\gamma_{sw}$ =1.0249 $gcc^{-1}$ at 20°C

Mass = $M_{sw}$ = 13.4715 g

Volume = $V_{sw}$ = 13.1442 cc

Free seawater mass ($M_{fsw}$) in pore space = ($M_{fsw}$)= (13.4715 g - 1.4639 g) = 12.0076 g

Total Seawater volume of hydration $V_{swh}$ = 1.4283 cc

Total Volume of Free SW pore space $V_{sw}$ = (13.1442 cc - 1.4283 cc) = 11.7159 cc

c. Clay Mineral solids Density = $\gamma_{cm}$ = 2.68 $gcc^{-1}$

Mass = $M_{cm}$ = 4.2717 g

Volume = $V_{cm}$ = 1.5939 cc

**B.** ***Calculation of Selected Mass Physical Properties***

1. (a) Total water content Corrected for Salinity. $\omega_{cs}$ = 223.46% (Dry Mass of Minerals + $OM_d$)

Uncorrected for hydrated OM

(b) Corrected free seawater content $\omega_{csomh} = (M_{fsw})/(M_{cm} + M_{omh})$ (6)

Free seawater is the water not bound to OM, i.e., the total seawater ($M_{sw}$) minus the water of OM hydration. See Eq. (3) and section VIII.E. Hydrated Organic Matter (above).

$\omega_{csomh}$ = [(13.4715 g - 1.4639 g)/ (4.2717g + 3.2207 g)] x 100 = 12.0076/7.4924 =

$\omega_{csomh}$ = 1.602637 x 100 = 160.2637%

This is the free seawater content (percent dry mass of clay minerals plus semisolid OM) corrected for hydrated organic matter. The differentiation of water content types provides new insight in the role of seawater in marine muds. Note: Some seawater is strongly bound as inter-layer water between clay domains that are stacked face-to-face in mud deposits. This inter-layer water is considered part of the free pore water in this paper and is not part of the water of OM hydration. The mass of this water strongly bound to the clay can be as large as a few tens of percent of the dry clay mass when the sediment contains a significant proportion of smectite. In highly consolidated mud deposits these types of water can diminish and virtually be non-existent following geological consolidation and high stress compaction processes that transforms mud to laminated shale. In addition, the total water content also includes the seawater attached to the surface of hydrated OM, the seawater attached to the surface of mineral solids, and if present the water attached to the surface of gas. Here these are considered as part of the free seawater.

2. Wet Bulk Density

$\gamma_t = M_T/V_T$ $M_T$ is Total Mass, $V_T$ is total Volume (7)

(13.4715 g SW + 4.2717 g CM + 1.7568 g $OM_d$) /(13.1442 cc $V_{sw}$ +1.5939 cc $V_{cm}$ +1.2116 cc $V_{omd}$) =

$\gamma_t$ = 19.5 g /15.9497 cc = 1.22259 $gcc^{-1}$

3. Porosity ($n_{cv}$): Ratio of the Volume of seawater in the free void space to the total sample Volume ($V_T$ = volume of clay mineral plus volume of hydrated organic matter plus Volume of the free seawater).

$$\text{Corrected Porosity} \quad n_{cv} = V_{fsw}/V_T \text{ at 100\% Saturation} \tag{8}$$

$V_{fsw}$ is the free flow seawater; it does not include water of hydration that is not available for free flow in pore space. Using corrected data for all three Phases the porosity at 100% saturation is calculated here:

$$n_{cv} = V_{fsw}/V_T = 11.7159 \text{ cc}/15.9497 \text{ cc} = 0.7345 \text{ or } 0.7345 \times 100 = 73.45\%$$

4. Void Ratio $e_{cv}$ = volume of the voids/volume of "solids" (solids CM + $OM_h$ hydrated). Volume of the voids ($V_v$) is actually the same $V_{fsw}$ as in Eq. (7), for 100% saturation.

The void space considered here is the free water volume at the given saturation of 100%. The volume of the solids is the volume of the clay minerals plus the volume of the $OM_h$.

$$\text{Corrected Void Ratio } e_{cv} = V_{fsw}/(V_{cm} + Vom_h) \tag{9}$$

Where $V_{cm}$ = Volume of minerals and $Vom_h$ = volume of hydrated $OM_h$

$Vom_h$ = 2.6399 cc, $V_{cm}$ = 1.5939 cc    $V_{fsw}$ = 11.7159 cc

$e_{cv}$ = 11.7159 cc/1.5939cc + 2.6399 cc = 11.7159/4.2338 = 2.7672

Note the porosity $n_{cv} = e_{cv}/(1 + e_{cv})$

**C. *Discussion of Fee Gas in Marine Mud Deposits***

In general gases dissolve in water to a greater extent the higher the pressure and the lower the temperature. All the gases of the atmosphere may be present and dissolved in the pore water of sediments as well as gases created by biological activity or geological processes. Dissolved gases may be released under laboratory conditions from sediment samples recovered from deep or cold sediments deposits in response to changes in pressure and temperature. This change from dissolved state to the more voluminous gas state has a potential to alter the fabric and mass physical properties of the recovered sediments (Imbert, 2012; Andreassen *et al.,* 2017).

Much more important both for its effects on the properties of recovered marine sediments and of sediments in situ, however, is the presence of free undissolved gases within the sediment deposit. At depths below a few meters of overlying water, only a few materials are commonly present as free gases in the pore water: methane and other low-molecular-weight hydrocarbons, carbon dioxide, and hydrogen sulfide. Of these, methane and its oxidation product carbon dioxide are by far the most common (Egger *et al.,* 2018). Methane is generated in situ by microbial-mediated decomposition of organic matter (biogenic process) and by breakdown of petroleum deposits when the ambient temperature at depth becomes high enough (thermodynamic process) (Arche, 2007). In addition, below water depths of about 300 meters, pressure and temperature conditions cause methane to form ice-like crystals of methane hydrate (Woods, 2002). As with free gas, this solid material alters the acoustic properties of the sediment. It may be present within the pore spaces and/or may cement the mineral grains together (Waite *et al.,* 2008; Gabitto and Tsouris, 2010; Marín−Moreno *et al.,* 2017). At atmospheric pressure, the methane hydrate decomposes, with every cubic centimeter forming ~0.8 cc of water and about 160 cc of methane gas and other trace gases. Hydrates with similar properties containing methane and smaller proportions of ethane or propane also are found in submarine sediments (Maekawa, 2001).

**1.** ***Consider a 100% Saturated mud and a Partially Saturated Marine Mud Sample***

We use the same sediment properties above, however, we include free gas ($V_{fg}$) when present, distributed throughout the void space. In this example we use the "Degree of Saturation" (S) of a partially saturated mud to be (S) = 90% of the voids. Historically this has been predicated on the concept that the total seawater after laboratory drying of a mud sample is all free pore water from the total void space (standard marine geotechnical practice, Lambe and Whitman, 1969). By definition, the degree of seawater saturation (S%) is expressed in percent as the ratio of the volume of the available free seawater ($V_{fsw}$) to the total volume of the void space $V_v$ Eq. (10). When the total free water volume is equal to the total void space ($V_{fsw} = V_v$) the percent saturation is equal to 100%.

$$S\% = V_{fsw}/V_v \times 100 \qquad (10)$$

In the case of the void ratio and porosity calculations of a mud with hydrated organic matter the total seawater is not all free seawater; the seawater of hydration ($V_{swh}$) is physico-chemically bound to the organic matter and is not free. The total seawater mass, corrected for salinity and obtained by oven drying, includes seawater of organic matter hydration plus the free seawater in the pore space.

An example of the degree of saturation at 100% is based on seawater hydrating organic matter and free seawater filling the void space in the mud. The calculations follow:

Total volume ($V_T$) of 3 Phases at 100% Saturation is $V_T$

$V_T$ = 2.6399 cc ($Vom_h$) + 11.7159 cc ($V_{fsw}$) + 1.5939 cc ($V_{cm}$) = 15.9497 cc

Total mass ($M_T$) of the 3 Phases at 100% Saturation = $M_T$

$M_T$ = 3.2207 g ($M_{omh}$) + 12.0076 g ($M_{fsw}$) + 4.2717 ($M_{cm}$) = 19.5000 g

An example of a partially saturated sediment with 10% of the void space occupied by free gas ($V_{fg}$) at laboratory temperature and pressure; therefore 10% of the total free pore water at 100% saturation volume is 1.17159 cc ($V_{fsw}$) and consequently the original total sample volume of 15.9497 cc has lost free seawater equivalent to the volume of the free gas ($V_{fg}$) or 11.7159 cc ($V_{fsw}$) - 1.17159 cc ($V_{fg}$) = 10.5443 cc of free seawater and 1.17159 cc of free gas. The volume of free gas may be distributed throughout the sample volume or be in one or more larger void volumes. The added volume of free gas and the loss of free water now changes the total wet bulk density of the sediment mass.

The total volume of a partially saturated ($V_{TPS}$) mud with four Phases considering each Phase is: Total Volume $V_{TPS} = V_{omh} + V_{fsw} + V_{cm} + V_{fg}$ (11)

$V_{TPS}$ = 2.6399 cc + 10.5443 cc + 1.5939 cc + 1.1716 cc = $V_{TPS}$ = 15.9497cc

The total mass of a partially saturated mud with four Phases considering each Phase is:

Total mass $M_{TPS} = M_{omh} + M_{fsw} + M_{cm} + M_g$ (12)

$M_{TPS}$ = 3.2207 g + 10.8068 g + 4.2717 g + 0.0015781 g = 18.30078 g Total Mass

Wet Bulk Density $\gamma_{twg}$ = 18.30078 g/15.9497 cc = 1.14741 $gcc^{-1}$ (with 10% gas)

Mass of gas = $M_g$ = 1.1716 cc x 0.001347 $gcc^{-1}$ = 0.0015781 g

If we average the mass of the two gases, methane and carbon dioxide, the most prevalent types of free gas in surficial sediments, we have: ~0.001347 $gcc^{-1}$ for the average of the two [0.717 $kg/m^3$ (.000717 $gcc^{-1}$), gas $CH_4$ (methane) and $CO_2$ 1.977 g/L (0.001977 $gcc^{-1}$)] at atm. pressure and $20^0$C.

The degree of seawater saturation is the ratio of the free water volume to the total volume of large and small void spaces. Corrections should be made for hydrated OM, when present, to correct for the available free water. Thus, for this example of a partially saturated mud:

$S_c\% = (V_{fsw}/V_v) \times 100$ (13)

The Volume of the free seawater ($V_{fsw}$) = 10.5443cc

The Volume of the total void space ($V_v$) = 11.7159cc

$S_c\%$ = 10.5443/11.7159 x 100 = 89.9999

The Degree of Saturation $S_c\%$ = 90.00% Saturation with 10.00% free gas

***2. Summary and Comparison of Saturation Calculations***

The Wet Bulk Density $\gamma_t$ = 1.22259 $gcc^{-1}$ at 100% Saturation

The Wet Bulk Density $\gamma_{twg}$ = 1.14741 $gcc^{-1}$ at 90% Saturation

Important to note is that with free gas present the porosity and void ratio would change from an in situ high-pressure at depth in the sediment to a lower pressure in the laboratory. The reader knows well that the gas present in situ would expand at laboratory pressure and temperature. The expansion and the compressibility of marine gas is variable and not all types of gases present in situ have the similar responses to pressure changes which complicates making volumetric predictions and calculations of the actual changes in gasses present from laboratory to in situ conditions without additional extensive laboratory analyses.

Regarding the laboratory measurement of the percent saturation in sediment samples, based upon the first author's experience working in various marine research laboratories, it is important to note that a common practice in assessing and determining the percentage of free gas in a sample was to use a water content measurement in proximity to a sample used for the determination of wet bulk density or a water content measurement following the drying of the material within the calibrated tube that was determined from a gravimetric and volumetric measurements of the sediment sample in a calibrated tube of 40-60 $cm^3$. Error could be introduced by assuming similar water contents in proximity to and outside the tube sample or that the water contents were uniform laterally and vertically throughout in the sample tube which is generally not the case in marine sediments. Variability in sediment physical properties over short distances is common (Bennett and Nelsen, 1983). For precise measurements of percent saturation, water content measurement should be made on the entire tube sample with measurement of water loss, total mass of the dry sediment, and with reliable calibrated measurements of tube volume and mass. With samples containing substantial amounts of OM, volumetric and gravimetric calculations measurements should be made of each sediment Phase (water, solids, OM, and free gas) including the contribution and correction for salinity of the seawater. The measurements of the Phases and bulk properties can be used to determine a total sample volume, mass, and the wet bulk density, porosity, void ratio, percent saturation, etc. of mud samples using a calibrated subsampling tube.

## X. ***Simulations of Fine-Grained Sediment Phases and Properties with Hydrated Organic Matter***

This section presents simulations of the four Phases (water, clay minerals, hydrated organic matter), with 100% saturation, for mass physical static and dynamic properties. To provide context for magnitude of the property changes due to hydrated OM, additional simulations are included for properties of sediment at 90% saturation. These simulations span a wide range in normally consolidated muds over a range of variables, including high porosity muds. "Variability in the physical properties of mud is the rule rather than the exception" (Bennett and Nelsen, 1983) and mud deposits are anisotropic and inhomogeneous over nanometer, centimeter, and meter scales. Limited lateral tortuosity measurements on the nano-meter scale of remolded marine clay show tortuosity ranges of 1.0 - 1.14 over porosity ranges of 80-100% (Douglas, 2014). The mass-volume of hydrated OM alters the interconnected pore volume in marine mud and likely increases the tortuosity of these interconnected pore volumes.

**A. *Simulation variables and range of values***

Under normal consolidation, water content decreases with depth below seafloor. However, marine mud deposits with low seawater content may have high or low %TOC values (Fig. 1). (Bennett *et al.,* 1985; Busch and Keller, 1981, 1982). Available data from locations in Fig. 5 indicate the most common %TOC values are less than 3 - 4%, however, values above ~9% do occur. Therefore, four separate ranges of total water content (percent dry mass) values are selected for four different ranges of %TOC as given in Table 2. Densities of clay minerals are given in Table 1. As previously noted in Section III.A, the mass of dry OM is computed from %TOC (via the Redfield ratio) and hydrated organic matter mass and volume can be computed.

Only particulate OM is considered in this simulation. Input parameters and value ranges used in simulations are given in Table 2. In these simulations, %TOC and water content are always used wth the full range of values. Here, the volume of OM water of hydration in the simulations is limited to a maximum of ~30% of the total pore water to permit simulations to apply to sediments containing a large fraction of smectite clay minerals. Dry and hydrated OM densities are not considered independent variables in these simulations, they are treated as independent pairs. For simplicity, no silt or sand components are included in these simulations and no limits are imposed on the amount OM relative to the amount of clay.

**Table 2. Simulation Values and Ranges**

| Parameters | Range | Parameter | Range |
|---|---|---|---|
| TOC % | 0.4 and ≤ 2 | Water Content | 30 – 210 % |
| TOC % | >2 and ≤ 5 | Water Content | 30 – 240 % |
| TOC % | >5 and ≤ 9 | Water Content | 200 – 330 % |
| TOC % | >9 and ≤ 13 | Water Content | 290 – 750 % |
| Clay Mineral Average Density g/cc | 2.45, 2.52, 2.60, 2.68, 2.75, 2.85 | (6 values) | |
| Seawater Density g/cc | 1.024 | constant value | |
| Dry OM Density g/cc<br>Hydrated OM Density g/cc | 1.40, 1.45, 1.50, 1.60<br>1.12, 1.22, 1.28, 1.35 | (four paired values) | |

Table 2. Input parameters and value ranges used in simulations. The water content is the percent with respect of the total amount of salt free water to the total dry mass in the simulated laboratory sample. See Calculation of Selected Properties: Sediment Phases, Section C. Seawater. %TOC is a common measure of total organic matter used to determine the organic matter mass. Seawater density is held constant throughout.

The relative span of values for each input parameter indicates it's relative importance to the calculated quantity of free pore water due to hydrated OM. Relative span was estimated by the (maximum value – minimum value) divided by the median value. Calculations of the relative span of important variables (Section VIII.) show that water content and %TOC are the principal determinants of OM hydration variability by a nearly factor of ten. OM (dry and hydrated) and clay mineral densities should not be ignored since the more clay, the more OM is generally observed (Furukawa, 1996), but the seawater density relative span, (1.020 -1.028) / (1.024) is less than 1%, a factor of 3 smaller than OM and clay density spans. Thus, seawater density is held constant herein.

Figure 6a shows plots of simulated total water content [Eq. (3)] versus wet bulk density (Fig. 6a). Usually, all of total water content is considered as free pore water which flows freely under any differential fluid pressure. Shown in Fig. 6b is a plot of wet bulk density versus water content that has been corrected for hydrated OM by treating the seawater of OM hydration as a component of OM not as free pore water. Since wet bulk density is the ratio of total mass to total volume, the differentiation of seawater as a combination of free pore water and seawater bound with OM to form a semisolid, does not change the total mass nor the total volume.

In calculations and comparisons involving wet bulk density it is tacitly accepted that all seawater moves freely under stress. The wet bulk density values associated with the best fit water content corrected for hydrated OM (Fig 6b) are noted here as "empirical" wet bulk density values. Fig. 6c indicates that, although the wet bulk density doesn't change due to hydrated OM, the impact of hydrated OM on water content does alter the empirical fit with wet bulk density. Fig 6.c shows the difference between the two best-fits (corrected and uncorrected water content for hydrated OM) at coincident water content values is ~1-3%. Empirical wet bulk density may fill a need for a characterization based on the reduced free pore water content due to hydrated OM for use with wet bulk density empirical classifiers, descriptors, or models (such as for consolidation, sound speed, and attenuation).

**Fig. 6 Simulations of wet bulk density for the water content, %TOC, hydrated OM, and clay density values in Table 2**

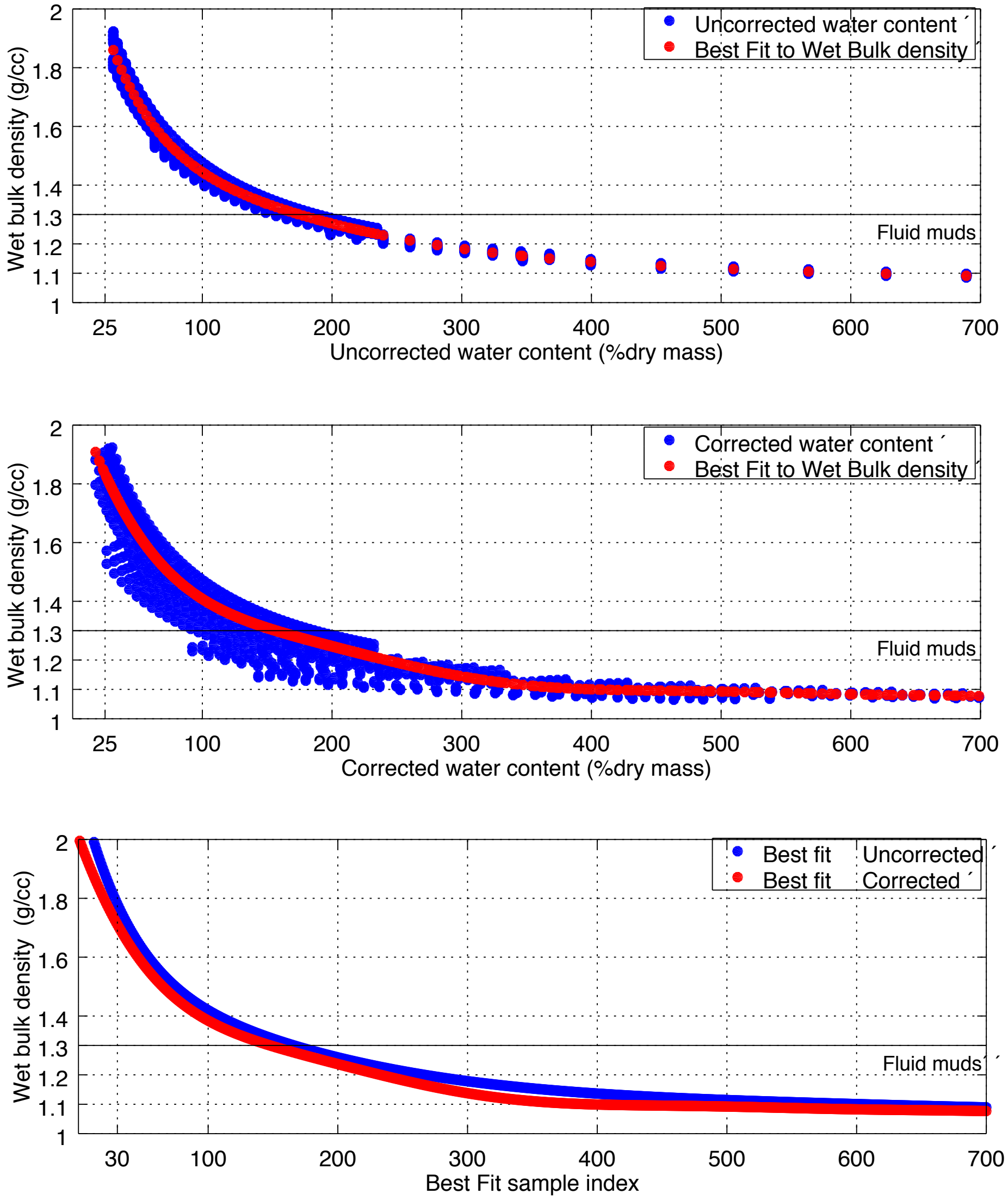

Figure 6. (a) Uncorrected water content treats all water as free pore water. The correction for hydrated OM (b) treats some water as bound water of OM hydration (see text). (c) The "empirical" impact on wet bulk density based upon the difference between the best-fit curves in (a) and (b).

Fig. 7 illustrates that the total water content components (free pore water and water bound to organic matter) are a source of variations in porosity. Water content is historically considered free pore water and porosity is used to empirically determine the consolidation as a function of depth in the sediment. If OM is present, accounting for hydrated OM may be necessary.

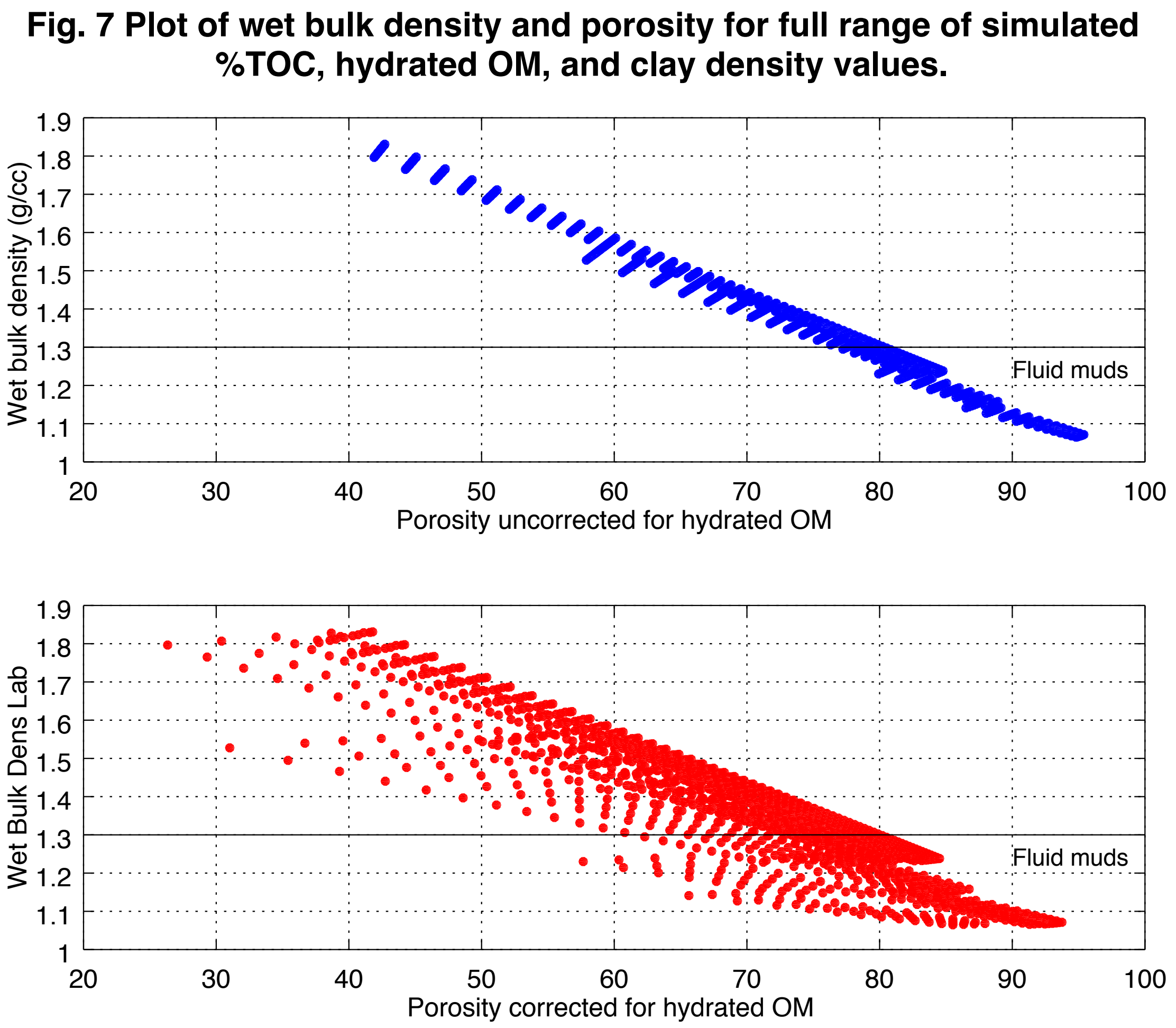

Figure 7. (a) Variations in simulated uncorrected porosity. (b) Larger variations in porosity due to corrections for hydrated OM (compared to water content in Fig. 6). Here meaningful variations may occur through the entire range of mud porosities.

**B.** ***Absolute Relative Differences***

Absolute relative differences are commonly used to evaluate the magnitude of the relative change to aid in decision making. It is given by

$$\%ARD = |(x_1 - x_2)| \div x_2 \times 100 \qquad (14)$$

Often, the value of $x_2$ is chosen to be a reference value (for calibration) or a mean value (for analysis) and is usually reported as a percent. Here, $x_1$ is the mass physical property value *uncorrected* for hydrated OM and $x_2$ is the same property value corrected for hydrated OM. Absolute relative differences are computed using varying combinations and permutations of the parameters in Table 2 to obtain both uncorrected and corrected values and to calculate %ARD for water content, wet bulk density, porosity, and void ratio to determine and compare the impact of hydrated OM. The %ARD values are sorted into 100 equally-spaced bins that span the entire range of differences. Thus, each bin represents a percentile and the number of differences in each bin is divided by the total number of differences to give the magnitude of each bin. Tables and plots of absolute relative differences (%ARD) will be used to compare the changes in bulk properties due to hydrated OM density. Table 3 (upper portion) summarizes the absolute relative differences (Eq.14) for selected mass-physical properties at selected percentiles of the %ARD distribution along with the mean and standard deviation. Simulations include all values and variables given in Table 2.

### C. *Saturation Differences Comparison*

To render context to the magnitude of the absolute relative differences between corrected and uncorrected for hydrated OM properties (Table 3), a comparison is made with a second set of relative differences that are more familiar to the reader. These are relative differences between a 90% void saturation and 100% void saturation (Table 3, lower portion). The saturation differences serve as a benchmark for the corresponding difference due to OM hydration correction. Here, 10% of the free pore water is replaced by gas (equal volumes of methane and $CO_2$) and all simulation values for both 90% and 100% saturations are uncorrected for hydrated OM. This yields a constant 10% difference between the water content of the two fluid saturations. Table 3 (lower) reports these absolute relative differences using 100% saturation as the reference value [Eq. 14].

**Table 3. Simulated effects of hydrated OM on geotechnical properties and comparison with a 10% void-space saturation effect.**

**%Absolute Relative difference between uncorrected and corrected properties for OM hydration.**

| Percentiles | 15th | 25th | 50th | 75th | 85th | 92nd | mean | std |
|---|---|---|---|---|---|---|---|---|
| Wet bulk density | 0 | 0 | 0 | 0 | 0 | 0 | 0 | 0 |
| Wet bulk density (empirical) | 0.4 | 0.6 | 1.3 | 1.9 | 2.4 | 3.7 | 1.7 | 1.8 |
| Water Content (% dry mass) | 3.2 | 4.6 | 9.7 | 21.5 | 30.7 | 46.2 | 18.1 | 23.1 |
| Porosity | 1.3 | 1.9 | 3.7 | 7.0 | 10.6 | 16.8 | 6.2 | 7.7 |
| Void ratio | 5.5 | 8.3 | 17.0 | 39.0 | 54.4 | 79.3 | 31.7 | 40.7 |

**%Absolute relative difference between 90% and 100% void-space saturation**

| Percentiles | 15th | 25th | 50th | 75th | 85th | 92nd | mean | std |
|---|---|---|---|---|---|---|---|---|
| Wet Bulk Density | 9.7 | 10.8 | 12.8 | 13.8 | 14.3 | 15.5 | 12.3 | 2.4 |
| Water Content (%dry mass) | 10.0 | 10.0 | 10.0 | 10.0 | 10.0 | 10.0 | 10.0 | 0.0 |
| Porosity (uncorrected for OM) | 6.3 | 6.8 | 7.5 | 7.8 | 8.1 | 8.5 | 7.3 | 0.9 |
| Void Ratio (uncorrected for OM) | 16.2 | 19.9 | 28.1 | 34.7 | 39.2 | 50.5 | 29.1 | 12.8 |

Table 3. Absolute relative differences (Eq. 14) are sorted into percentiles for each property. Selected percentiles are given for comparison. For OM corrected properties, the water of hydration for OM is not treated as free pore water but is bound by OM. The lower table gives absolute relative differences between simulations with two different void saturations, 100% saturation (no gas) and a 90% saturation (10% gas saturation in the void space). No hydrated OM correction is made for either of the two saturations used to compute absolute relative differences in the lower table (see text).

In the saturation percentiles, the wet bulk density saturation differences increase from ~9 to ~14% over the span of the water content. These range to more than 4 times higher than "empirical" wet bulk density differences due to hydrated OM. Water content (percent dry mass) differences due to OM hydration range from 4 - 22% (over the 25th - 75th percentiles) compared to constant 10% for saturation differences. The mean OM hydration difference for water content is ~18%, nearly twice that of the saturation difference. For porosity, the OM hydration mean absolute difference is ~6% with a std. dev. of ~8%. The saturation mean absolute difference is ~7% and std. dev. is 1%. Although the means are nearly equal, the standard deviations are quite different. Void ratio is the most variable property for both sets of differences. For the saturation differences the 50th percentile and the means are similar (~28% and ~29%); for the OM hydration differences the two values are ~17% and 32% and the standard deviation is two times higher than for the saturation.

The partial gas saturations percentiles are more nearly Gaussian. The non-Gaussian nature of the percentiles for hydrated OM differences (Table 3) are evidenced by the ratio of the standard deviation to the mean. The non-Gaussian nature of the absolute relative differences are due to the hydrated OM's dependence on two components, free pore water volume and total OM mass (see Eq. 4). One could conclude if a 10% saturation difference would be important to a process or phenomena under study, then the hydrated OM correction for free pore water may also induce significant differences for at least some combinations %TOC and water content.

**D.** ***Comparison of Absolute Relative Differences due to Hydrated OM.***

Fluid muds are defined by their wet bulk density value (Fig. 1) and this allows unambiguous delineation of the simulated mud properties into fluid-mud and semi-solid mud classes. Here fluid mud indicates both fluid mud and fluid-like mud (Fig. 1) and semi-solid mud denotes mud densities > 1.30 g/cc and the condition that acoustic shear wave propagation is supported. Fluid muds are important to scientific and engineering disciplines especially for coastal and tidally dominated depositional environments as well as along continental slopes. Semi-solid muds include plastic mud, and plastic-like mud (Fig. 1) and are delineated by consolidation properties not density. Due to overburden pressure, muds consolidate and mud density increases, squeezing out the free seawater and causing clay fabric re-orientation towards to dominantly face-of-face. At higher consolidation levels, the overburden pressure begins to squeeze out the seawater of OM hydration (Fig. 1).

For these simulations (Table 2), semi-solid mud is contained within %TOC values below 6% while fluid mud spans almost the entire %TOC range. Table 4 summarizes the absolute relative differences (%ARD) simulation results for semi-solid and fluid mud classes. There are significant differences due to hydrated OM between these two mud classes. Semi-solid mud void ratio %ARD is lower by roughly a factor of two than for fluid-mud and at higher percentiles it is lower by a larger factor. Fluid-mud exhibits higher %ARD percentile differences than does semi-solid mud for both water content and void ratio properties in all percentiles. The mean and standard deviation in fluid mud void ratio differences are more than a factor of 2 higher than values for semi-solid muds. Porosity differences between the two mud classes are nearly similar. Empirical wet bulk density is not considered because it is based on best fits to the simulations.

The IQR (inner quartile range) value is defined as the difference between the 25th and 75th percentile values and is a descriptor of likely relative difference one might expect with real data. Statistical methods are used for examining the simulation results to compare and convey the magnitude and range of hydrated OM's impact on mass physical properties of marine mud over a wide range of variables and values. However, these simulations are not random processes and the "outliers" are not outliers in the statistical sense. Here they are a direct consequence of linearly increasing OM mass (given by %TOC) and the discreet but unequally spaced OM dry and hydrated densities chosen for simulation (Table 2). The variability of low and mid-range (equally spaced) hydrated OM densities is generally indicated by the IQR box and high OM density by the “outliers”.

**D.1.** ***Semi-solid Mud - Relative Differences due to hydrated OM***

Figure 8 shows box plots for the absolute relative difference (%ARD, Eq. 14) between uncorrected and corrected for hydrated OM mass physical properties for semi-sold mud. Box plots are shown for several spans of %TOC values and the %ARD scale (y-axis) varies differently for each property. Porosity, void ratio, and water content each show linear increasing trends of relative differences with %TOC for both median and IQR. This slope appears small because of the large range of values in the %ARD differences to accommodate “outliers” (due to highest clay density in T able 2). The minimum void ratio relative difference exceeds 5~7% for almost any water content when the *%TOC is above ~1%*. The %ARD for empirical wet bulk density difference (bottom right) is based on the best-fit of curves of uncorrected and corrected (for hydrated OM) simulations (Fig 6c). Visual comparison of these curves suggests that the difference between two best-fit smoothing operations decreases with water content. Hydrated OM impact is highly dependent on the on the mass-physical property under consideration. As previously noted, semi-solid mud spans %TOC values up to ~6%. Semi-solid muds share the same water content values with fluid-muds only the range ~90-200 percent dry mass (Fig. 6a). Thus, mud type comparisons necessarily encompass both similar and different subsets of the simulation parameters.

*Table 4.  Comparison of variability in Absolute Relative differences between fluid-mud and semi-solid mud simulations.*

| | Fluid mud | | | Semi-solid mud | | |
|---|---|---|---|---|---|---|
| | **IQR** | **Coefficient of variation** | **Coefficient of dispersion** | **IQR** | **Coefficient of variation** | **Coefficient of dispersion** |
| Wet bulk density(empirical) | 1.1 | **1.29** | **0.52** | **1.2** | 0.89 | 0.40 |
| Water Content (% dry mass) | **24.7** | 1.14 | **0.69** | 10.6 | **1.21** | 0.56 |
| Porosity | 4.4 | 1.05 | 0.54 | **5.60** | **1.32** | **0.60** |
| Void ratio | **44.4** | 1.1 | **0.67** | 16.9 | **1.16** | 0.53 |

Table 4. As noted, semi-solid mud spans only %TOC values up to ~6%. Only the the range  ~100-200 (percent dry mass) do semi-solid muds share the same water content values with fluid-muds. Thus, these mud types are mostly different subsets of the values used in the simulations. Note: the bold values indicate the higher value between the mud types.

**Fig. 8 Box plot comparison of absolute relative differences for semi-solid mud simulations of bulk properties due to hydrated OM corrections.**

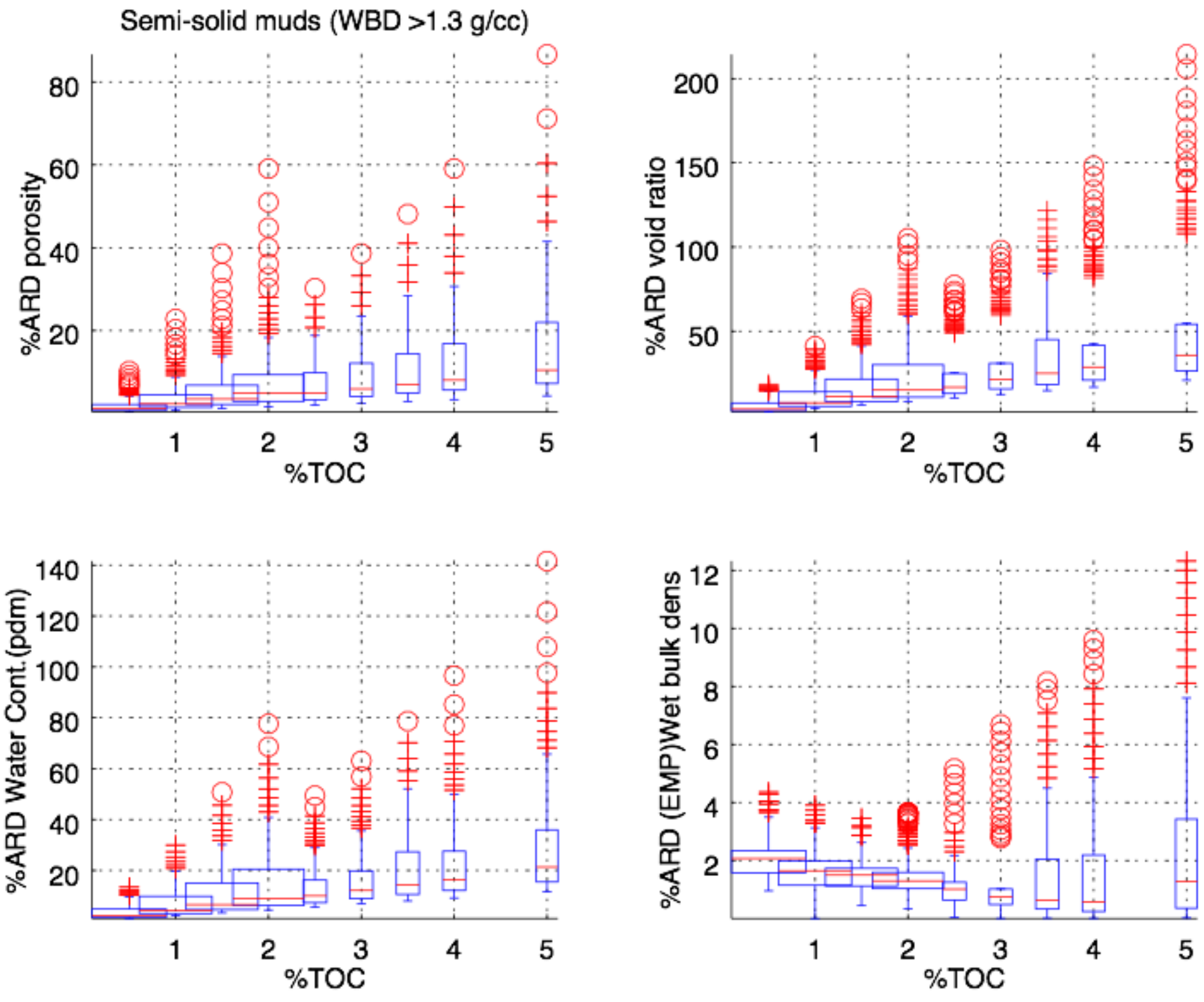

Figure 8. Inter Quartile Range comparison of absolute relative differences (%Ard) as a factor of %TOC for semi-solid mud simulations. IQR is indicated by the height of the blue box (Table 4) and the median is indicated by the horizontal red line. Vertical axis scales for %Ard are unique to each subplot. The IQR loosely corresponds to lower and medium values of OM density (Table 2) and the "outliers" to the highest value of hydrated OM density,

Table 4 compares the variations in %ARD between fluid and semi-solid muds. The dispersion measure is used to compare the differences in two groups under different conditions. IQR is a useful measure but uses only the center half the percentile differences. Variation measure uses the standard deviation and the mean while dispersion uses the sum and differences of the 25th and 50th percentiles.

Measures for some quantities between the two muds are nearly equal but overall fluid mud has slightly higher dispersion for %ARD than semi-solid mud for all properties except porosity. Note that semi-sold mud exhibits higher variations for porosity %ARD for all three measures. For both mud types, void ratio has the largest %ARD and the highest IQR but not the highest variation. Thus, %ARD determination from average conditions or "like" regions my not be reliable under some conditions. For void ratio and water content (pdm), the IQR ( centroid of the %ARD distribution) is higher for for fluid mud. For empirical wet bulk density the IQR is slightly larger for semi-solid mud but the variation and dispersion measures are higher for fluid mud.

These %ARD measures of variability are high enough for both fluid and semi-solid mud to indicate that the four components that determine the hydrated OM phase (seawater volume of OM hydration, clay density, dry OM density, and fully hydrated OM density) cannot be ignored if highly accurate estimates of porosity, void ratio, and free pore water are required. In addition, hydrated OM corrections are required under most real-world conditions (set forth in Table 2), sometimes even when OM hydrated mass is relatively small. Additional laboratory analysis and computations are required as dictated in Section 2.

**D.2.** ***Geoacoustic Relevance***

Empirical relationships between sediment mass-physical static and dynamic properties and acoustic properties as a function of frequency have been used for decades to predict and model acoustic transmission (Hamilton 1971 and 1972). Historically clay muds were identified as one of two classes based on grain size (either silty mud or clayey mud). Acoustic shear speeds in clay muds are typically less than ~100 meters/sec for semi-solid muds and decrease as the porosity increases. Solid components of clay fabric and semi-solid hydrated OM generate and support shear waves at velocities dependent upon clay-fabric stiffness. Sediment classification compilations (Jackson and Richardson, 2007) show considerable shear velocity variability with porosity (uncorrected for hydrated OM).

For marine clays, the cited acoustic shear speed data (Figure 5.11 of Chotiros, 2017) varies from ~45 m/s to ~1 m/s over the (uncorrected) porosity range from ~65% to ~85%. The clay data trend is clearly distinct from the sand and silt data (which is quasi-linear). The expected curve for marine clay shear wave velocity is concave downward and the cited shear wave speed data was fit to separate porosity curves for fresh water and seawater saturated clays. No difference in curvature due to salinity was identified. The difference in porosity between the two curves is ~6-9% and increases slightly with increasing porosity.

In Fig. 8, (the %ARD porosity box-plot for semi-solid mud) spans %TOC values from 1-5%. The corresponding porosity absolute relative differences due to hydrated OM range from 1~10% for the porosity box plot median value . Thus, hydrated OM may well be responsible for the observed shear wave velocity variability. Trial data which includes corrections for hydrated OM mud samples is needed for verification. Well-documented properties of hydrated OM provide both physical foundations for the modifications which Chotiros (2021) reports to be required to repurpose the extended Biot model for sands and silts to include mud. These foundations are (1) hydrated OM is present as particulates in the pore fluid and (2) OM binds a significant fraction of the pore fluid.

***D.3. Compression Index for semi-solid mud***

The highest absolute relative differences between corrected and uncorrected for hydrated OM are for void ratio. Void ratio plays a determining role in sediment compaction, marine mud elasticity and plasticity, and other engineering properties (Terzaghi and Peck, 1967). The compression index is defined as the change in void ratio divided by the change in the effective vertical stress (kPa). Hydrated OM can inhibit in-situ dewatering under lower stress conditions and high stress conditions can squeeze out the seawater of OM hydration. Compression tests with numerous samples provide data for empirical curves using varying parameters (usually void ratio, water content, or liquid limit). Models using only void ratio (uncorrected for hydrated OM) to quantify the compression index value for various types of clay muds are available for multiple mud types (Bahrami and Marandi, 2018), They use the form y= m*x + b where y is the compression index, x is the void ratio, and m and b are empirical constants determined by best-fits to measurements. To quantify the effect of hydrated OM on such an empirical model requires a trial that includes a best-fit determination of m and b based the void ratio corrected for hydrated OM.

## XI. *SUMM*ARY

The major purpose of this research was to consider the presence and importance of the mass and volume of hydrated organic matter ($OM_h$) in mud deposits. Marine mud deposits are the most frequently occurring sediment types in ocean environments from shallow to deep seawater and consists fundamentally of seawater and siliciclastic clay mineral solids often with some sand and silt size particles of different mineralogy and free seawater. However, many mud deposits contain two (2) additional Phases that include hydrated organic matter and/or free gas. This paper demonstrates that hydrated organic matter, when present, is an important mud Phase. Historically, in many disciplines the total dry organic carbon (TOC) was reported in percent mass of the dry solids and neglected reporting the mass of hydrated organic matter. The complex variability and chemistry of mud Phases, including the clay microstructure determine the mass physical properties of mud deposits on the scale of centimeters to meters. This variability drives the static and dynamic stress-strain behavior of marine fine-grained mud deposits and consolidation processes.

Although further research is needed, the authors conclude that only a fraction of the total water content is constrained within small clay aggregates that can contain some hydrated OM. Historically, the hydration of the organic matter has not been considered by many disciplines and the organic matter on a dry basis has been generally reported as total organic carbon, without recognition and quantification of hydrated OM.

This research paper considers hydrated OM as a semi-solid Phase and the seawater of hydration is physico-chemically bound in organic matter forming a semi-solid mass-volume. When hydrated OM is not considered, the calculated void ratios, porosities, and free water contents are based on a higher mass and volume of free pore water than is present. Thus, errors in the calculations and predictions of the mass-physical properties increase as the ratio of the volume of OM hydration to the total pore water volume increases and as the ratio of the hydrated OM mass to clay-mineral mass increases. Subsequently, estimates of shear strength, consolidation behavior, and dynamic object penetration-depth will be in error. Understanding the consolidation process of marine mud deposits can be poor even when a moderate amount of hydrated OM is present. The bearing capacity of mud decreases significantly with the presence of significant amounts of hydrated OM and total water content.

Simulation results of mass physical properties reveal that over the total water content range of 120-360% dry mass, hydrated OM produces discernible and important variability in the wet bulk density – water content curve (Fig. 6). Each mass-physical property exhibits a unique span of relative differences. Porosity differences for corrected and uncorrected calculations of hydrated OM are highly variable on the order of 2-15% (Fig. 7) and the void ratio differences are on the order of ~10–60% (Table 4). Small amounts of hydrated OM can yield large differences in bulk properties and large amounts of hydrated OM may yield lower differences depending upon the hydrated-OM to dry-clay mass ratio and the hydrated-OM to free-pore-water volume ratio. Variability is to be expected in empirical shear-speed vs porosity curve. Bulk property differences due to hydrated OM at 100% saturation were mostly comparable to bulk property differences due to a 10% change in saturation (Table 3), but the OM hydration differences are more widely distributed. The authors believe that researchers presently engaged in various fields of marine and riverine mud research and multiple areas of engineering will benefit from carefully considering the impacts of hydrated OM in muds.

The presence of hydrated organic matter in fine-grained mud deposits (riverine, coastal, and shallow and deep oceanic water) would be an important research topic in terms of the bulk modulus and compressibility of various amounts and types of hydrated OM. Future studies can be focused on different salinity water, types of degraded organic material, types of clay minerals, and variability of the density of the four sediment Phases that comprise many mud deposits. The compressibility of material Phases is the inverse of the bulk modulus k or B = 1/k and the data that is provided here should improve the understanding of the role of hydrated OM in the correction of several mass physical and dynamic properties of marine mud. The research results discussed in this paper can have an impact in the fields of marine geology, biogeochemistry, geoacoustic, geotechnology and geotechnique and can help in understanding of the impact burial of dynamic objects penetrating a deposit and the settlement of submarines and other objects placed in a deposit or on the surface of marine mud.

## XII. *ACKNOWLEDGEMENTS*

Dr. Allan Pierce and Dr. William Siegmann graciously provided their knowledge and encouragement, through technical discussions, and correspondence on various topics in this paper. The authors appreciate the invitations to present topics discussed in this paper at two ASA Meetings by Dr. Mohsen Badiey, at University of Delaware and Dr. Megan Ballard, at University of Texas. Dr. Richard Faas (deceased) provided insightful discussions regarding fluid mud properties over many years of cooperative research on marine mud. We are also grateful to Dr. Nicholas Chotiros for insights and comments. We appreciate the research by Conrad Curry for compiling the information on the %TOC in marine and riverine sediment for several locations around globe.